\documentclass{article}

\PassOptionsToPackage{numbers, compress}{natbib}
\usepackage[preprint]{neurips_2026}

\usepackage[utf8]{inputenc} 
\usepackage[T1]{fontenc}    
\usepackage{hyperref}       
\usepackage{url}            
\usepackage{booktabs}       
\usepackage{amsfonts}       
\usepackage{nicefrac}       
\usepackage{microtype}      
\usepackage{xcolor}         

\usepackage{graphicx}
\graphicspath{{figures/}}
\newcommand{\code}[1]{\texttt{\detokenize{#1}}}

\title{A General B\'ezier Tree Encoding Counterfactual Framework for Retinal-Vessel-Mediated Disease Analysis}

\author{%
  Tan Su\textsuperscript{1}\quad
  Ethan Elio Meidinger\textsuperscript{2}\quad
  Lin Gu\textsuperscript{3,$*$}\quad
  Ruogu Fang\textsuperscript{4,$*$}\\[2pt]
  \textsuperscript{1}\,Department of Electronic and Electrical Engineering, Southern University of Science and Technology\\
  \textsuperscript{2}\,School of Data Science,University of Virginia, Charlottesville\\
  \textsuperscript{3}\,Research Institute of Electrical Communication, Tohoku University\\
  \textsuperscript{4}\,J. Crayton Pruitt Family Department of Biomedical Engineering, University of Florida\\[2pt]
  \texttt{12311316@mail.sustech.edu.cn, vtt4vx@virginia.edu,}\\
  \texttt{lin@tohoku.ac.jp, ruogu.fang@bme.ufl.edu}\\[2pt]
  \textsuperscript{$*$}Corresponding authors.
}

\begin{document}

\maketitle

\begin{abstract}
The geometry of the retinal vessel is a key biomarker of vascular diseases, yet clinical evidence remains primarily observational. Existing generative counterfactuals intervene only at the image-level disease label, failing to isolate explicit anatomical structure. To address this limitation, we propose the B\'ezier Tree Encoding Counterfactual Framework (BTECF). By abstracting vascular networks into interconnected cubic-B\'ezier segments, BTECF establishes a disease-agnostic representation in which structural topology is explicitly preserved and atomically perturbable. Coupling this encoding with a diffusion-based generator enables parameter-level \emph{do}-interventions on explicit geometric axes (e.g., tortuosity, caliber) while preserving background fundus textures. We validate BTECF on diabetic retinopathy, together with independent cohorts for ischemic stroke and Alzheimer's disease. Isolated counterfactual interventions produce dose-responsive shifts in classifier predictions; a matched pixel-drop control attenuates this response by an order of magnitude or more, ruling out out-of-distribution generation artifacts. By enforcing causal isolation between vessel topology and pixel-level confounders, BTECF provides a unified generative paradigm for hypothesis verification across systemic diseases. To support reproducibility, the code will be publicly released upon acceptance.

\end{abstract}

\section{Introduction}
\label{sec:intro}

\begin{figure}[t]
  \centering
  \includegraphics[width=\linewidth]{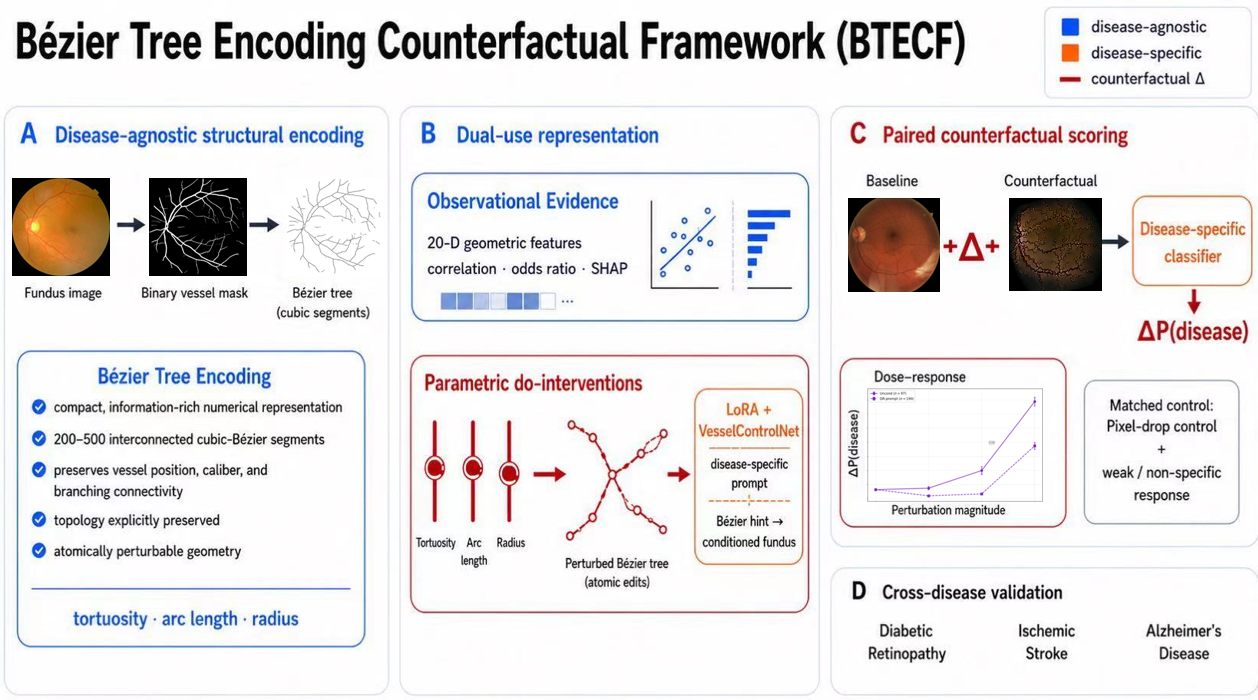}
        \caption{BTECF overview. Color coding: blue = disease-agnostic, orange = disease-specific, red = counterfactual response $\Delta$. \textbf{(A)} Cubic-B\'ezier encoding of the segmented fundus. \textbf{(B)} Dual-use representation: $20$-D geometric features for observational analysis (top); B\'ezier hint conditions a LoRA + VesselControlNet generator for parametric \emph{do}-interventions (bottom). \textbf{(C)} Paired counterfactual scoring yields $\Delta P(\mathrm{disease})$; matched pixel-drop control isolates the geometric driver. \textbf{(D)} Cross-disease validation on diabetic retinopathy, ischemic stroke, and Alzheimer's disease.}

  \label{fig:framework-formalization}
\end{figure}

The retina provides a unique window for direct, non-invasive optical observation of the central nervous system (CNS)~\citep{Zhao2024EyeBrain}. Clinical studies have demonstrated that retinal microvascular alterations are associated with the onset of stroke~\citep{wong2001retinal,mcgeechan2009prediction,doubal2010fractal}, cardiovascular events and hypertension risk~\citep{wong2004retinal,sandoval2021retinal}, as well as Alzheimer's disease and cognitive decline~\citep{frost2013retinal,lemmens2020systematic,Zhang2022RetinaAD}. In particular, morphological changes in the retinal vasculature, including caliber variation, tortuosity, and branching topology, serve as key indicators of systemic disease pathology~\citep{arnould2023using,yusufu2025retinal,doubal2010fractal}.

Consequently, retinal vessel segmentation has long been a central research focus. From handcrafted morphological operators~\citep{hoover2000locating,soares2006retinal} to deep convolutional neural networks (CNNs)~\citep{Ronneberger2015UNet}, and more recently Transformer-based architectures~\citep{zhou2022automorph,Chen2024TransUNet}, segmentation performance has been steadily improved, approaching practical limits. 

However, despite recent advances enabling the prediction of Diabetic Retinopathy \citep{liu2025stmfdrnet}, Stroke \citep{khalafi2025stroke}, and Alzheimer's Disease \citep{jamshidiha2025transformer} from retinal images, most existing approaches rely on Transformer architectures that encode the entire image holistically, without explicitly modeling the underlying vascular structure. This limitation prevents the formulation of counterfactual and interventional queries, such as: “If we intervene on a specific geometric feature X of a given vessel, how would the disease prediction change?”

To address this limitation, we introduce B\'ezier Tree Encoding~\citep{bezier1972numerical}, an information-rich numerical representation of the retinal vascular tree, analogous to the SWC parametric standard for neuronal morphology~\citep{mehta2023online}. By abstracting a $512 \times 512$ pixel mask into a compact set of $200$--$500$ interconnected cubic-B\'ezier segments, this encoding maps vessel position, caliber, and branching connectivity directly to clinically interpretable parameters. Because the structural topology is explicitly preserved, the three geometric axes (tortuosity, arc length, and radius) become atomically perturbable, providing the technical prerequisite for parameter-level \emph{do}-operator interventions. The encoding is dual-use: it yields a 20-dimensional feature vector for observational analysis and a spatial hint for generative conditioning. Building on this encoding, we propose the B\'ezier Tree Encoding Counterfactual Framework (BTECF) (Figure~\ref{fig:framework-formalization}). Unlike pixel-mask conditioning approaches~\citep{feng2024diversified,go2024generation,guo2024controllable} or classical parametric vessel renderers~\citep{bonaldi2016automatic,castro2020visual}, BTECF couples the parametric encoding with a diffusion-based generator to synthesize single-feature, decoupled counterfactual samples.

BTECF operates through three modules tied to the panels of Figure~\ref{fig:framework-formalization}. The disease-agnostic upstream (Fig.~\ref{fig:framework-formalization}A, blue) extracts a B\'ezier Tree Encoding from segmented fundi~\citep{zhou2022automorph}, transforming a vessel mask into a parametric tree with three atomically perturbable axes (tortuosity, arc length, radius). This encoding is dual-use (Fig.~\ref{fig:framework-formalization}B): a $20$-dimensional geometric feature vector supports observational analysis (top, disease-agnostic), while the same tree, rasterized as a three-channel B\'ezier hint, conditions a VesselControlNet~\citep{zhang2023adding} plugged into a LoRA-adapted~\citep{hu2022lora} diffusion generator with disease-specific text prompts that synthesize photorealistic counterfactual fundi from parametric \emph{do}-interventions on the three axes (bottom, disease-specific). The intervention effect is then quantified by a within-start paired counterfactual scoring protocol (Fig.~\ref{fig:framework-formalization}C): a disease-specific classifier scores each baseline against its perturbed counterfactual to produce $\Delta P(\mathrm{disease})$, while a matched pixel-drop control isolates the geometric driver from generator-side artifacts. The same pipeline is instantiated on diabetic retinopathy, ischemic stroke, and Alzheimer's disease (Fig.~\ref{fig:framework-formalization}D).

We validate BTECF primarily on diabetic retinopathy (DR), alongside independent cohorts for ischemic stroke and Alzheimer's disease to test cross-disease portability. Observational analyses on the B\'ezier Tree Encoding feature vector align with established epidemiological and histopathological mechanisms via SHAP top-5 mappings. Isolated counterfactual interventions on the three geometric axes produce dose-responsive shifts in classifier predictions; a matched pixel-drop control attenuates this response by over an order of magnitude, ruling out out-of-distribution generation artifacts. BTECF therefore provides mechanistically interpretable ``observation-intervention'' dual-modal evidence, offering a unified paradigm for hypothesis verification across systemic diseases. Nevertheless, our current validation is primarily computational and prospective clinical follow-up is needed.

\paragraph{Our contributions}
\begin{enumerate}

\item \textbf{B\'ezier Tree Encoding Counterfactual Framework (BTECF).} We couple an automatically extracted, atomically perturbable cubic-B\'ezier tree with a diffusion prior. This dual-use encoding enables both observational analysis and parameter-level \emph{do}-interventions on vascular topology, preserving background fundus textures.

\item \textbf{Parameter-level counterfactual verification.} Isolated interventions on three geometric axes produce dose-responsive prediction shifts across three systemic diseases (DR, stroke, AD); a matched pixel-drop control rules out generative artifacts.

\item \textbf{Identification of language-prior confounding.} We identify a vision--language decoupling effect: textual disease prompts can mask classifier sensitivity to pure geometric signals, exposing a confounder for multimodal medical AI pipelines.
\end{enumerate}


\section{Related Work}
\label{sec:related}

\paragraph{Observational vessel-geometry studies}
Cohort studies across multiple retinal-vessel-mediated diseases share three methodological limitations. They are correlation-only, unable to answer the interventional ``what if we change $X$'' question that our framework targets. Cohort sizes are often limited to hundreds to low-thousands of participants in DR-specific vessel-geometry studies~\citep{cheung2008retinal,klein2004relation,velayutham2020extended,xu2026quantitative}, and although stroke- and dementia-specific cohorts are larger, they report narrow vessel-geometry dimensionality. Measurement relies on semi-automated pixel-skeleton tools (most commonly IVAN with the Parr--Hubbard CRAE / CRVE formula), restricting each study to a small set of univariate summaries such as central retinal vessel-equivalent caliber (CRAE/CRVE), fractal dimension and mean tortuosity, without resolving the full caliber distribution or branching topology. Our 20-dimensional B\'ezier feature space resolves observational signal more finely. Mainstream DR observational studies report four characteristic geometric changes: vessel arc length decreases as the vasculature atrophies~\citep{forster2021retinal,pramil2021macular}; fractal dimension declines~\citep{broe2014retinal,popovic2019fractal,forster2021retinal,pramil2021macular}; vessel tortuosity rises~\citep{sasongko2011retinal,sasongko2016retinal,forster2021retinal,fathimah2025retinal,xu2026quantitative}; and vessel caliber follows a biphasic trajectory across DR severity, widening in early DR and narrowing in late stages~\citep{ashraf2021retinal,cheung2008retinal,klein2004relation,velayutham2020extended,xu2026quantitative}. These observations anchor the three perturbation axes (tortuosity, arc length, radius) we select in \S\ref{sec:perturb-axes}.

\paragraph{Generative and counterfactual fundus imaging }
Mask-conditioned deep generators synthesize retinal images from rasterized anatomical maps, including vessel masks and DR lesion masks~\citep{feng2024diversified,go2024generation,guo2024controllable}. However, raster-space conditioning does not expose clean parameter-level handles on vessel caliber, arc length, or topology. Classical parametric vessel-tree renderers~\citep{bonaldi2016automatic,castro2020visual} expose interpretable geometry but operate without a photorealistic texture prior, limiting their use as inputs to image-trained classifiers. On the causal side, deep structural causal models~\citep{pawlowski2020deep} target medical imaging in general, while the most relevant fundus counterfactual~\citep{ilanchezian2025development} achieves lesion-level realism via disease-label intervention, but the authors note that vessel-geometry changes are poorly captured by their model. CauDR~\citep{wei2024caudr} applies do-operations on frequency-domain features for DR domain generalization rather than on anatomically interpretable parametric vessel geometry, and counterfactual contrastive learning~\citep{roschewitz2025robust} performs counterfactual generation on acquisition-domain factors (e.g., scanner) for representation robustness rather than on parametric vessel geometry. An adjacent line of work uses quadratic-B\'ezier curves for procedural vessel synthesis as segmentation pretraining augmentation~\citep{niu2023bezier}, with synthetic curves randomly generated and lacking pathological grounding. In contrast to these approaches, BTECF performs do-interventions directly on anatomically interpretable parametric vessel geometry and measures the downstream disease response.

\section{Method}
\label{sec:method}

\subsection{Framework Overview}
\label{sec:framework-formalization}

BTECF is organized as six modules: vessel segmentation~\citep{zhou2022automorph}, B\'ezier Tree Encoding parameterization, a LoRA~\citep{hu2022lora} domain adapter on Stable Diffusion 2.1~\citep{rombach2022high}, a VesselControlNet~\citep{zhang2023adding} conditioned on the B\'ezier hint, a downstream disease-specific classifier, and a within-start paired counterfactual scoring protocol. Vessel segmentation, B\'ezier parameterization, and the scoring protocol are disease-agnostic and transfer with zero retraining; LoRA, ControlNet, and classifier are instantiated per disease. The B\'ezier parametric pipeline yields both a 20-dimensional geometric feature vector for the observational analysis and a three-channel ControlNet hint for the generative and counterfactual analyses. Figure~\ref{fig:framework-formalization} shows the full pipeline; per-module deployment cost is in Appendix~\ref{app:modules}.

\subsection{B\'ezier Parametric Hint}
\label{sec:bezier-hint}

Cubic-B\'ezier curves~\citep{bezier1972numerical} are a standard parametric basis in computer graphics for smooth curve representation. Given a segmented vessel mask, we split the skeleton at branch nodes and fit each resulting polyline piecewise into 200--500 cubic-B\'ezier segments per fundus, preserving topological continuity at branch nodes and geometric accuracy along each segment. This parametric representation is then rasterized into a three-channel ControlNet hint: Channel~0 carries the distance-transform radius field of the raw mask, Channel~1 renders the B\'ezier curves at variable radius read from Channel~0, and Channel~2 is a Gaussian-smoothed version of Channel~1. This three-channel split enables the decoupled per-axis counterfactual interventions of \S\ref{sec:perturb-axes}. Per-segment equations, rasterization parameters, the empirical Channel-0 invariance check, and anatomical-coherence properties are in Appendix~\ref{app:hint}. Figure~\ref{fig:hint-decomposition} visualizes the B\'ezier fit and the resulting three-channel hint on a representative fundus.

\subsection{Training and Inference}
\label{sec:training}

We finetune Stable Diffusion 2.1~\citep{rombach2022high} with a LoRA adapter~\citep{hu2022lora} (rank~32, 36k steps) and \code{noise_offset = 0.1}, motivated by prior observations that offset noise improves diffusion models' ability to represent global brightness shifts. Empirically, in our fundus finetuning runs, it reduced gray/blue color drift. A VesselControlNet~\citep{zhang2023adding} checkpoint is trained on our B\'ezier hint and used by counterfactual intervention. For discriminative evaluation features, we fine-tune a BiomedCLIP encoder~\citep{zhang2023biomedclip} on the disease labels of the target cohort. The downstream classifier is trained from scratch on the leak-free DR cohort: EfficientNet-B2 serves as the primary scorer for the counterfactual intervention, ResNet-50 provides the cross-backbone sensitivity check, and a ViT-B/16 trained at the same scale underperforms at this training budget and is treated as diagnostic rather than confirmatory. Per-backbone AUC and sensitivity figures are in Appendix~\ref{app:classifier}.

At inference we use classifier-free guidance~\citep{ho2022classifier} with weight $w = 7.5$,
\begin{equation}
  \epsilon_{\mathrm{final}} = \epsilon_{\mathrm{unc}} + w \cdot (\epsilon_{\mathrm{cond}} - \epsilon_{\mathrm{unc}}),
  \label{eq:cfg}
\end{equation}
where the unconditional branch corresponds to an empty text string and is exactly the branch we analyse under the Uncond condition (\S\ref{sec:cf-uncond}). For the counterfactual intervention, the same starting grade-0 fundi are generated under two conditions: (A) DR prompt and (B) Uncond, with all perturbation configurations identical across conditions, so that the (A) vs.\ (B) comparison isolates the prompt's effect. The reported effect is the within-start paired difference
\begin{equation}
  \Delta_i \;=\; P(\mathrm{disease} \mid \mathrm{pert}, \mathrm{start}_i) \;-\; P(\mathrm{disease} \mid \mathrm{baseline}, \mathrm{start}_i),
  \label{eq:delta}
\end{equation}
aggregated over the $n$ valid starts as $\mathrm{mean} \pm \mathrm{SEM}$; any systematic generator bias enters both sides as a common term and cancels in $\Delta$. 95\% confidence intervals use the one-sample $t$-distribution and $p$-values test $\Delta = 0$. Per-backbone hyperparameters, the prompt-pairing protocol, the empirical color root-cause analysis, and inference hyperparameters (DDIM steps, ControlNet residual strength, LoRA weights, latent offset) are in Appendix~\ref{app:training} and~\ref{app:inference}.

\subsection{Counterfactual Perturbation}
\label{sec:perturb-axes}

The counterfactual intervention perturbs three B\'ezier feature axes (tortuosity, arc length, and radius), selected on four grounds: (i) observational analysis (SHAP top-5 features, \S\ref{sec:obs}); (ii) atomic perturbability under the parametric representation; (iii) cross-disease literature correspondence~\citep{forster2021retinal,frost2013retinal,lemmens2020systematic,sandoval2021retinal,doubal2010fractal,wong2001retinal,klein2004relation,cheung2008retinal,velayutham2020extended,xu2026quantitative,wong2004retinal}; and (iv) dose-response quantifiability (graded strengths per axis). Per-axis justification is in Appendix~\ref{app:axes}; the same three-axis selection applies to the cross-disease cases without re-selection.

Each axis is perturbed by a decoupled mechanism on the parametric representation. The \emph{tortuosity family} at multipliers $\alpha \in \{1, 2, 4\}$ perpendicularly displaces the interior control points $P_1, P_2$ by an amount proportional to the segment's chord length. For a cubic B\'ezier segment with chord vector $\vec{v} = P_3 - P_0$ of length $L = \|\vec{v}\|$ and unit perpendicular $\hat{n} = (-v_y, v_x)/L$, with a per-segment random sign $s \in \{-1, +1\}$ drawn uniformly,
\begin{equation}
  P_1' = P_1 + s \cdot \gamma \alpha L \cdot \hat{n}, \qquad
  P_2' = P_2 - s \cdot \gamma \alpha L \cdot \hat{n},
  \label{eq:tortuosity-perturb}
\end{equation}
where $\gamma = 0.15$ is the base displacement coefficient. The endpoints $P_0, P_3$ are preserved (chord length and topological connectivity invariant by construction); the antisymmetric $P_1, P_2$ assignment produces an S-shaped local curvature within each segment. The $\alpha = 1$ multiplier is therefore non-zero displacement; the zero-displacement reference is the baseline configuration that bypasses this function entirely. The \emph{arc-drop family} at strengths $10\%, 20\%, 30\%$ randomly deletes a fraction of B\'ezier segments, leaving the remaining segments intact. The \emph{radius family} at scaling factors $0.85, 0.70, 0.55$ multiplies the distance-transform field by a global factor, reducing caliber uniformly. A matched \emph{pixel-drop control} at the same three strengths deletes mask pixels uniformly at random and re-runs skeletonization and B\'ezier fitting on the degraded mask; it is deliberately not geometrically decoupled and serves as a non-decoupled null matching pixel-loss magnitude.

\subsection{Evaluation Protocols}
\label{sec:eval-protocols}

\paragraph{Observational analysis}
For each fundus in the DR cohort we automatically extract the 20-dimensional B\'ezier feature vector and apply three complementary statistical analyses: Spearman rank correlation captures the marginal monotonic trend; logistic regression yields per-SD and quintile (Q5/Q1) odds ratios that resolve linear and non-linear effects; TreeSHAP importance on a trained XGBoost model also reports a per-feature \emph{feature-versus-SHAP correlation} (\texttt{feat-SHAP corr}), defined as the Pearson correlation between each feature's raw value and its tree-conditional SHAP contribution. A sign disagreement between Spearman~$\rho$ and \texttt{feat-SHAP corr} is the statistical fingerprint of a biphasic, non-monotonic feature--risk relationship in which both extremes of the feature distribution associate with elevated risk: the marginal monotonic test flattens such a U-shape toward zero, while tree-based SHAP resolves the two arms through interactions with correlated features.

\paragraph{Discriminative benchmark}
An XGBoost benchmark in the main analysis cohort isolates how much disease discriminatory signal each feature representation carries. We compare the 20-dimensional B\'ezier geometric features with the frozen BiomedCLIP, both at native dimensionality and in a 20-dimensional PCA regime matched.

\paragraph{Counterfactual intervention}
\label{sec:cf-design}
We draw 500 grade-0 fundi from the held-out test split and apply 13 perturbation configurations (one baseline plus three doses each of arc-drop, radius scaling, tortuosity, and the pixel-drop null control), giving 6{,}500 ControlNet generations per prompt condition. All images are scored with the EfficientNet-B2 classifier (primary); a ResNet-50 sensitivity check is reported on the original subset $n = 50$. We report two complementary estimators: a \emph{population-level ATE} on all $n = 500$ starts, and a \emph{strict reverse-counterfactual subset} of starts with baseline DR probability below $0.3$ that pass a visual-fidelity filter (full thresholds in Appendix~\ref{app:inference}). The two are deliberately complementary: the population ATE provides high-power confirmation that the effect is not specific to a hand-picked subset, while the strict subset removes ceiling-effect starts and isolates the upper bound of disease-emergence response on negative-baseline starts. The Strict-subset yields are $n = 97$ (Uncond) and $n = 190$ (DR-prompt).

\section{Experiments}
\label{sec:experiments}

\subsection{Data and Setup}
\label{sec:data-setup}

\paragraph{DR primary case}
We use the Eyepacs, Aptos, Messidor Diabetic Retinopathy dataset on Kaggle, which curates four publicly available DR datasets into a single $600\times600$ resized collection (EyePACS~\citep{diabetic-retinopathy-detection}, APTOS-2019~\citep{aptos2019-blindness-detection}, APTOS Gaussian Filtered, and Messidor-2~\citep{decenciere2014feedback}). The redistribution applies manual augmentation to expand the dataset by approximately $55\%$, yielding $143{,}669$ images. The grades follow the 5-step ICDR clinical scale: $0 = $ no DR, $1 = $ mild NPDR, $2 = $ moderate NPDR, $3 = $ severe NPDR, $4 = $ proliferative DR. We use binary labels throughout: grade 0 as normal and grades 1--4 as DR. We enforce strict deduplication, removing 2{,}954 redundant rows to obtain the final splits of $114{,}209 / 13{,}269 / 13{,}237$ (train / val / test). The val + test union (\textbf{$n = 26{,}506$}) is the cohort of observational and discriminative analysis. Detailed deduplication procedure in Appendix~\ref{app:deduplication} and data license in
Appendix~\ref{app:assetslink}.

\paragraph{Cross-disease cases}
Two independent UK Biobank \citep{sudlow2015ukbiobank} cohorts with disjoint patient groups: an ischemic-stroke cohort (\textbf{$n = 11{,}361$} participants; $3{,}409$ evaluation, $12\%$ positive) and an Alzheimer's disease cohort (\textbf{$n = 1{,}877$} participants; $550$ evaluation, $7.5\%$ positive). Splits, leak-free deduplication, and disease-specific retraining are described in \S\ref{sec:cross-disease} and \S\ref{sec:framework-formalization}.

\subsection{Observational Findings}
\label{sec:obs}

The SHAP top-5 features on the DR cohort (Table~\ref{tab:track1-top5}) recover the geometric directions reported in mainstream DR observational and histopathological literature (\S\ref{sec:related}): arc length and radius decrease with disease, tortuosity rises, and branching density carries a non-linear (biphasic) signature. The complete 20-feature analysis is in Appendix~\ref{app:track1-full} and a detailed pathology analysis is in Appendix~\ref{app:dr-pathology}.

\begin{table}[!htbp]
  \centering
  \caption{SHAP top-5 features on the DR cohort. Metrics defined in \S\ref{sec:eval-protocols}. Significance: $***$ $p < 0.001$, $*$ $p < 0.05$. }
  \label{tab:track1-top5}
  \resizebox{\linewidth}{!}{%
  \begin{tabular}{clccccll}
    \toprule
    Rank & Feature & Spearman $\rho$ & feat-SHAP corr & OR/SD & Q5/Q1 OR & SHAP abs & Direction \\
    \midrule
    1 & \code{total_arc_length}    & $-0.358$*** & $-0.94$ & 0.55 & 0.20 & 0.456 & $\downarrow$ DR \\
    2 & \code{branching_density}   & $+0.014$*   & $-0.55$ & 1.21 & 1.09 & 0.364 & SHAP non-linear \\
    3 & \code{mean_radius}         & $-0.084$*** & $-0.90$ & 0.85 & 0.68 & 0.285 & $\downarrow$ DR \\
    4 & \code{mean_chord_length}   & $-0.079$*** & $-0.84$ & 0.80 & 0.67 & 0.231 & $\downarrow$ DR \\
    5 & \code{mean_tortuosity}     & $+0.086$*** & $+0.17$ & 1.00 & 1.60 & 0.160 & $\uparrow$ DR \\
    \bottomrule
  \end{tabular}%
  }
\end{table}

\subsection{Discriminative Results}
\label{sec:disc}

The 20-dimensional B\'ezier geometric features alone reach test AUC $0.706$, covering $61\%$ of the random-to-BiomedCLIP attainable gap; in a matched 20-dimensional regime the two feature sets attain AUC $0.71$ (B\'ezier) and $0.80$ (BiomedCLIP PCA) (Table~\ref{tab:track2}). The geometric parameters the counterfactual intervention intervenes on therefore carry principal disease-discriminative signal rather than auxiliary signal, so parameter-level perturbations have a mechanistic rationale for causal efficacy.

\begin{table}[!htbp]
  \centering
  \caption{XGBoost discriminative comparison on the DR cohort.}
  \label{tab:track2}
  \begin{tabular}{lccc}
    \toprule
    Feature set & Dim & Test AUC & 5-fold CV AUC \\
    \midrule
    B\'ezier geometric (this work) & 20 & 0.7058 & $0.7146 \pm 0.0037$ \\
    BiomedCLIP frozen features & 512 & 0.8358 & $0.8587 \pm 0.0031$ \\
    BiomedCLIP PCA to 20-d (EVR $= 85\%$) & 20 & 0.8028 & $0.8128 \pm 0.0030$ \\
    \bottomrule
  \end{tabular}
\end{table}

\subsection{Parameter-Level Counterfactual Intervention}
\label{sec:cf}

The complete results on the DR case follow (\S\ref{sec:cf-uncond}--\S\ref{sec:cf-guidance}); cross-disease validation is in \S\ref{sec:cross-disease}.

\subsubsection{Primary Result}
\label{sec:cf-uncond}

Empty-prompt generation (the unconditioned CFG branch) isolates vessel-geometry causation from text-priming artifacts. Under Uncond condition in the strongest tortuosity dose (\code{tortuosity_4x}), the EfficientNet-B2 classifier produces a population-level ATE of $\Delta = +0.432$ ($95\%$ CI $[+0.399, +0.464]$, $p < 10^{-10}$, $n = 500$); in the strict reverse-counterfactual subset ($n = 97$), the mean probability of DR increases from $0.064$ at baseline to $0.689$ after intervention ($\Delta = +0.625$, $p < 10^{-10}$). The matched pixel-drop control attenuates by $\sim\!25\times$. Under the matched DR-prompt condition, the same classifier produces a population-level ATE of $\Delta = +0.079$ ($95\%$ CI $[+0.036, +0.122]$, $p = 3.7 \times 10^{-4}$, $n = 500$); in the strict reverse-counterfactual subset ($n = 190$), the mean probability of DR increases from $0.064$ at baseline to $0.375$ after intervention ($\Delta = +0.311$, $p < 10^{-10}$). The DR-prompt ATE is therefore $\sim\!5.5\times$ smaller than the Uncond ATE at the population level and $\sim\!2.0\times$ smaller in the strict subset; this prompt-condition attenuation is analyzed in \S\ref{sec:vision-language}. The complete landscape of 13-configurations $\Delta$ (Table~\ref{tab:track4-full}) confirms tortuosity dominance; per-perturbation analysis and the causal-response cross-prompt curves are in Appendix~\ref{app:track4-full-table}.

\begin{table}[!htbp]                                                          \centering                                                                  \caption{Full 13-configuration cross-condition $\Delta$ table on the DR primary case (paired within-start differences vs.\ baseline). (A) DR-prompt at $n = 190$; (B) Uncond at $n = 97$; both at $w = 7.5$, EfficientNet-B2 scorer, strict reverse-counterfactual subset.}                                 \label{tab:track4-full}                                                     \begin{tabular}{lcc}                                                          \toprule                                                                    Perturbation & (A) DR-prompt $\Delta$ & (B) Uncond $\Delta$ \\              \midrule                                                                    \code{tortuosity_4x} (strong) & $+0.311$ ($p{<}10^{-10}$) & $+0.625$ ($p{<}10^{-10}$) \\                                                              \code{tortuosity_2x} (mid) & $-0.031$ & $+0.136$ \\                         \code{tortuosity_1x} (weak, 15\% chord) & $-0.044$ & $+0.011$ \\            \code{arc_drop_30} & $+0.002$ & $-0.011$ \\                                 \code{arc_drop_20} & $+0.025$ & $-0.009$ \\                                 \code{arc_drop_10} & $+0.017$ & $-0.014$ \\                                 \code{radius_x0.55} (strong thin) & $-0.059$ & $-0.020$ \\                  \code{radius_x0.70} (mid thin) & $-0.058$ & $-0.009$ \\                     \code{radius_x0.85} (weak thin) & $-0.058$ & $-0.026$ \\                    \code{pixdrop_30} (control) & $-0.044$ & $-0.036$ \\                        \code{pixdrop_20} & $-0.031$ & $-0.037$ \\                                  \code{pixdrop_10} & $-0.016$ & $-0.010$ \\                                  baseline & 0 (by def.) & 0 \\                                               \bottomrule                                                               \end{tabular}                                                            \end{table}         

\subsubsection{Backbone and Prompt Sensitivity}
\label{sec:cf-sensitivity}

The strongest tortuosity effect (\code{tortuosity_4x}) is positive and statistically significant in all four cells of the $2 \times 2$ backbone $\times$ prompt design (Table~\ref{tab:cf-2x2}); EfficientNet-B2 cells use the $n = 500$ strict subset; ResNet-50 cells use the $n = 50$ pilot subset. Therefore, the causal effect at the parameter-level reproduces between both downstream classifiers and both prompt conditions, indicating that the geometric driver is classifier-agnostic. ViT-B/16 trained on the same scale yields only one qualifying start and is excluded.

\begin{table}[!htbp]
  \centering
  \caption{Counterfactual response $\Delta$ on \code{tortuosity_4x} across the $2 \times 2$ backbone $\times$ prompt design. All four cells significant at $p < 5 \times 10^{-3}$.}
  \label{tab:cf-2x2}
  \begin{tabular}{lcc}
    \toprule
    Backbone & DR-prompt $\Delta$ & Uncond $\Delta$ \\
    \midrule
    EfficientNet-B2 & $+0.311$ & $+0.625$ \\
    ResNet-50       & $+0.294$ & $+0.377$ \\
    \bottomrule
  \end{tabular}
\end{table}

\subsubsection{Guidance-Weight Ablation}
\label{sec:cf-guidance}

We sweep the CFG guidance weight $w \in \{3, 5, 7.5, 10\}$ on the $n = 50$ pilot subset under the DR-prompt condition (Table~\ref{tab:guidance-ablation}). DR-prompt $\Delta$ rises monotonically across $w \geq 5$, recovering $15\%$, $54\%$, and $72\%$ of the Uncond ceiling $\Delta = +0.625$ (\S\ref{sec:cf-uncond}) at $w = 5$, $7.5$, $10$. This pattern shows that prompt and hint compete for the same classifier-relevant visual content (\S\ref{sec:vision-language}): as $w$ rises, the prompt provides a positive hint-aligned contribution that partially recovers the geometric signal, but never reaches the strength of the uncontested pure-hint condition.

\begin{table}[!htbp]
  \centering
  \caption{Guidance-weight ablation on \code{tortuosity_4x} ($n = 50$ pilot, DR-prompt). DR-prompt $\Delta$ partially recovers toward the Uncond ceiling as $w$ rises, but never crosses it. Uncond ceiling $\Delta = +0.625$ from the $n = 500$ main strict subset (\S\ref{sec:cf-uncond}); $w$ has no formal effect under \code{prompt_mode=uncond} since both CFG branches encode the empty string.}
  \label{tab:guidance-ablation}
  \begin{tabular}{clc}
    \toprule
    $w$ & DR-prompt $\Delta$ ($n$, $p$) & DR-prompt $\Delta$ / Uncond ceiling \\
    \midrule
    3.0 & 0.150 ($n{=}14$, $p{=}0.14$, n.s.) & 0.24 (n.s.) \\
    5.0 & 0.095 ($n{=}14$, $p{=}0.05$) & 0.15 \\
    7.5 & $0.340$ ($n{=}17$, $p{=}3.3{\times}10^{-3}$) & 0.54 \\
    10.0 & $0.450$ ($n{=}13$, $p{=}5.3{\times}10^{-5}$) & 0.72 \\
    \bottomrule
  \end{tabular}
\end{table}

\subsection{Cross-Disease Validation: Ischemic Stroke and Alzheimer's Disease}
\label{sec:cross-disease}

We apply BTECF to two UK Biobank cohorts spanning the vascular--neurodegenerative spectrum: ischemic stroke and Alzheimer's disease. The three-disease summary is in Appendix~\ref{app:cross-disease-summary} (Table~\ref{tab:three-disease}).

\paragraph{Ischemic stroke}
Spearman analysis on the 20-dimensional B\'ezier features yields nine significant features ($p < 0.05$), six of which align directionally with prior observations~\citep{wong2001retinal,mcgeechan2009prediction,doubal2010fractal}. Applying the same parameter-level intervention (\code{tortuosity_4x}) on a stroke-retrained EfficientNet-B2 classifier reproduces the monotonic dose-response: mean stroke probability rises from a baseline of $0.001$ to $0.002$ at \code{tortuosity_1x}, $0.018$ at \code{tortuosity_2x}, and $0.338$ at \code{tortuosity_4x} ($\Delta = +0.337$, $95\%$ CI $[+0.19, +0.49]$, $n = 17$, $p < 10^{-3}$). The matched pixel-drop control yields no significant effect (\code{pixdrop_30} mean $0.002$, $\sim\!200\times$ weaker than \code{tortuosity_4x}). Full stroke tables and cohort-design discussion are in Appendix~\ref{app:cross-disease-full}.

\paragraph{Alzheimer's disease}
The discriminative ranking inverts relative to DR (B\'ezier test AUC $0.646 >$ BiomedCLIP $0.454$), supporting that the AD signal is captured by parametric vessel geometry rather than by the lesion-pathological visual content. Univariate Spearman is power-limited at this cohort scale, but \code{std_curvature} reaches significance with the largest Q5/Q1 odds ratio across the three diseases ($3.84$), consistent with fundus-based AD work reporting retinal vascular biomarker differences in Alzheimer’s disease~\citep{frost2013retinal}. The tortuosity intervention reproduces the positive direction ($\Delta = +0.19$ on \code{tortuosity_4x}, $n = 30$); specificity to topological geometry is reduced relative to DR and stroke because the AD classifier's higher false-positive baseline (sensitivity $0.50$) lets the pixel-drop control also produce elevated probabilities (Appendix~\ref{app:ad-full}).


\section{Discussion}
\label{sec:discussion}

\label{sec:vision-language}                                                                                                         BTECF integrates a fully automated upstream that extracts cubic-B\'ezier control points from raw fundus images with a hybrid parametric-to-pixel counterfactual generator. The pipeline admits atomic do-interventions on individual geometric axes while yielding photorealistic counterfactual fundi at population scale. On the DR primary case, interventions on tortuosity, arc length, and radius produce dose-response curves aligned with prior observational and pathological evidence; the matched pixel-drop control attenuates this effect by an order of magnitude or more, ruling out generation-side artifacts. The same intervention reproduces across two re-trained classifiers and two re-trained ControlNets on ischemic stroke and Alzheimer's disease, supporting framework portability rather than a DR-specific artifact. The parametric do-operation also probes either arm of a biphasic distribution separately, decoupling cross-sectional confounding that observational analyses cannot. The experiments further expose a vision--language decoupling phenomenon: text prompts and the parametric hint occupy partially overlapping classifier-relevant visual channels, so the textural channel can mask the topological signal that downstream classifiers would otherwise leverage. We expect this confounder to generalize to other multimodal medical imaging pipelines.

\paragraph{Limitations}
Our framework's primary validation remains computational; prospective OCT-A or fluorescein-angiography follow-up is the most decisive next step, ideally paired with a blinded clinical evaluation of generated high-tortuosity fundi against real images. The strict reverse-counterfactual subset prioritizes causal purity over statistical power, and the AD cohort's small size leaves its classifier with a high false-positive baseline that weakens pixel-drop-control specificity (Appendix~\ref{app:ad-full}). The framework occupies the L2 parametric-to-pixel tier of an L0--L3 conditioning spectrum (Appendix~\ref{app:l0-l3}); L3 (a parameter-native ControlNet consuming B\'ezier control points as a token sequence) remains a future direction alongside representation-space CKA analyses of the vision--language decoupling mechanism and morphological distribution matching against real high-tortuosity cohorts.

\paragraph{Broader Impacts}
Beyond research use, parameter-level counterfactual imaging can audit deployed disease classifiers for geometry-driven failure modes and inform clinical-research interventions such as anti-VEGF therapy-response prediction. Conversely, the same generative capability could be misused for adversarial classifier probing, and model-level causal evidence risks being misread as biological proof in clinical contexts where prospective validation has not yet been performed.

\paragraph{Data and Reproducibility}
The Kaggle redistribution Eyepacs, Aptos, Messidor Diabetic Retinopathy (Appendix~\ref{app:assetslink}) is used for DR; UK Biobank cohorts (Appendix~\ref{app:assetslink}) are used for stroke and Alzheimer's disease. Code, model weights, and evaluation scripts will be publicly released upon acceptance. The total computation is approximately $100$ GPU-hours per disease on RTX-5090-class hardware.


\section{Conclusion}
\label{sec:conclusion}

We introduced the B\'ezier Tree Encoding Counterfactual Framework (BTECF), which establishes parameter-level counterfactual reasoning over retinal vessel geometry by bridging cubic-B\'ezier parametric encoding with diffusion-based ControlNet generation. Across diabetic retinopathy and two cross-disease cohorts (ischemic stroke and Alzheimer's disease), parametric do-interventions reproduce dose-monotonic shifts in disease prediction and isolate vessel topology as a causal driver of classifier output rather than an out-of-distribution artifact. The framework also exposes vision--language decoupling as a transferable confounder pattern for separating structural and textural channels in multimodal medical AI. Parameter-native control (L3) and prospective clinical validation are the natural next steps for advancing from model-level to biological causality.

\bibliographystyle{plainnat}
\bibliography{references}

\appendix

\section*{Technical Appendices and Supplementary Material}

\section{Framework Modules and Cross-Disease Deployment Recipe}
\label{app:modules}

The six modules of BTECF are summarized in Table~\ref{tab:framework-modules}. Modules~1, 2, and~6 (vessel segmentation, B\'ezier parameterization, counterfactual scoring) are disease-agnostic and transfer with zero retraining: they reuse the released code and weights without modification. Modules~3, 4, and~5 (LoRA, ControlNet, classifier) are disease-specific and require retraining on the new disease's fundus cohort. LoRA and ControlNet must be co-trained in the same batch with matched \code{noise_offset = 0.1}; mixing an old LoRA with a freshly trained ControlNet produces persistent grayscale or blue-white color-drift artifacts at inference. The observational analysis depends only on the disease-agnostic modules and therefore requires no retraining; the counterfactual intervention (parameter-level causality) requires the full six-module deployment.

\begin{table}[!htbp]
  \centering
  \footnotesize
  \caption{The six modules of BTECF, grouped by cross-disease transferability. Modules~1, 2, and~6 transfer with zero retraining; modules~3, 4, and~5 require per-disease retraining.}
  \label{tab:framework-modules}
  \resizebox{\linewidth}{!}{%
  \begin{tabular}{@{}clp{4.4cm}p{2.8cm}p{3.4cm}@{}}
    \toprule
    \# & Module & Role & Training-domain dependency & Cross-disease transferability \\
    \midrule
    1 & AutoMorph vessel segmentation~\citep{zhou2022automorph} & Fundus to binary vessel mask & General vessel segmentation & Disease-agnostic, zero retrain; OOD risk on new cameras \\
    2 & B\'ezier parameterization & Mask to control points, geometric features, three-channel raster hint & No learned parameters & Disease-agnostic, zero retrain \\
    3 & LoRA domain adapter on SD 2.1 & Adapts the diffusion backbone to the cohort visual style & Target-cohort fundus images (no labels required) & Per-cohort retrain when camera or acquisition shifts \\
    4 & VesselControlNet (B\'ezier-conditioned) & B\'ezier hint to disease-conditioned fundus & Mask, label, and fundus  & Per-disease retrain; the core counterfactual probe \\
    5 & Downstream disease classifier & Scores counterfactually generated fundi & Fundus and label pairs & Per-disease retrain \\
    6 & Counterfactual scoring protocol & Within-start paired $\Delta$ and statistics & No learned parameters & Disease-agnostic \\
    \bottomrule
  \end{tabular}}
\end{table}

\section{Implementation Details}

\subsection{B\'ezier Hint Pipeline}
\label{app:hint}

Given a binary mask $M \in \{0, 1\}^{H \times W}$, the hint is constructed in three steps.

\paragraph{Skeletonization}
Morphological skeletonization produces a 1-pixel-wide topological skeleton, discarding caliber while preserving connectivity.

\begin{figure}[!htbp]
  \centering
  \includegraphics[width=\linewidth]{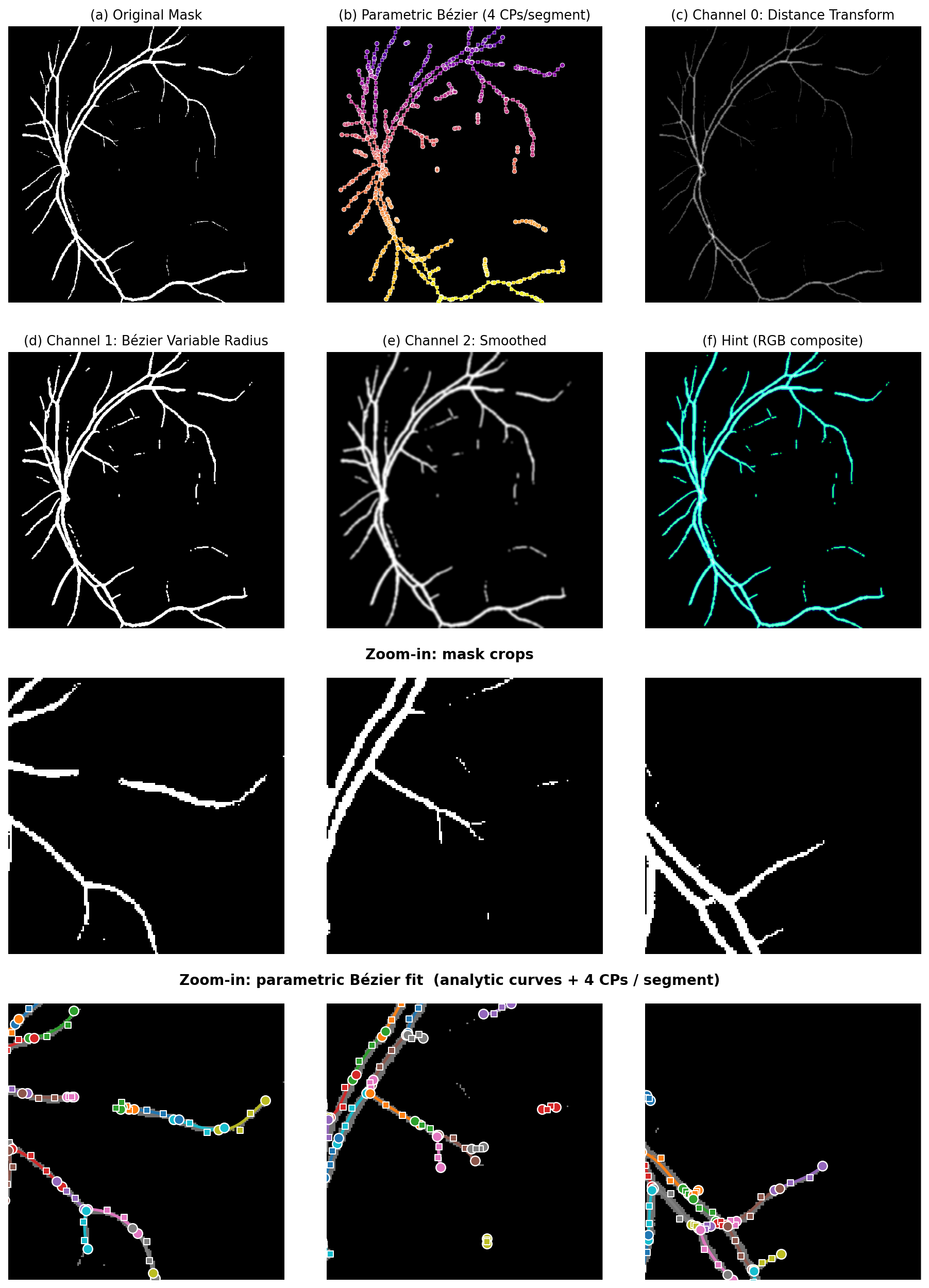}
  \caption{Decomposition of the three-channel B\'ezier hint derived from the binary vessel mask. \textbf{Top section, panels (a)--(f):} Channel~0 is the distance-transform radius field, Channel~1 is the B\'ezier variable-radius render, and Channel~2 is the Gaussian-smoothed version of Channel~1. The three channels are stacked into the final RGB hint passed to ControlNet. \textbf{Bottom two rows:} zoom-in examples cropped from the same source mask, illustrating per-segment fitting. The mask-crop row shows three local regions of the binary mask; the parametric-fit row overlays the cubic-B\'ezier reconstruction on the same crops, with each color denoting one fitted segment. Within a single color, the four nearest markers (two circles for endpoints $P_0, P_3$ and two squares for interior control points $P_1, P_2$) form the four control points of one minimum-unit cubic B\'ezier.}
  \label{fig:hint-decomposition}
\end{figure}

\paragraph{B\'ezier fitting}
The skeleton is first broken at degree-1 endpoints and degree-$\geq 3$ branch nodes, producing a set of ordered open polylines (typically 10--30 per fundus). Each polyline is then chopped into overlapping chunks of 30 ordered skeleton points (4-point overlap between adjacent chunks for endpoint continuity), and each chunk is fit by least squares to a cubic B\'ezier with arc-length parameterization within the chunk:
\begin{equation}
  B(t) = (1 - t)^3 P_0 + 3(1 - t)^2 t P_1 + 3(1 - t) t^2 P_2 + t^3 P_3, \qquad t \in [0, 1],
\end{equation}
with $(P_0, P_3)$ clamped to the chunk endpoints and $(P_1, P_2)$ the interior control points fitted by least squares. A typical fundus produces between 200 and 500 such cubic segments. Per-segment fitting examples are shown in the bottom two rows of Figure~\ref{fig:hint-decomposition}.

\paragraph{Fusion rendering into three channels (Figure~\ref{fig:hint-decomposition})}
Channel~0 is the Euclidean distance-transform radius field of the raw mask, normalized to $[0, 1]$. Channel~1 is the B\'ezier variable-radius render: each B\'ezier segment is sampled adaptively at twice the polyline chord length (with a floor of $20$ samples per segment, yielding sub-pixel coverage at $512 \times 512$), and each sample point is dilated by the local radius read from the distance-transform field. Channel~2 is a Gaussian-smoothed version of Channel~1 with kernel size $7 \times 7$ and standard deviation $\sigma = 2.0$. The three channels are stacked into an RGB hint of resolution $512 \times 512$ and rescaled to $[-1, 1]$ before being fed to ControlNet.

\begin{figure}[!htbp]
  \centering
  \includegraphics[width=\linewidth]{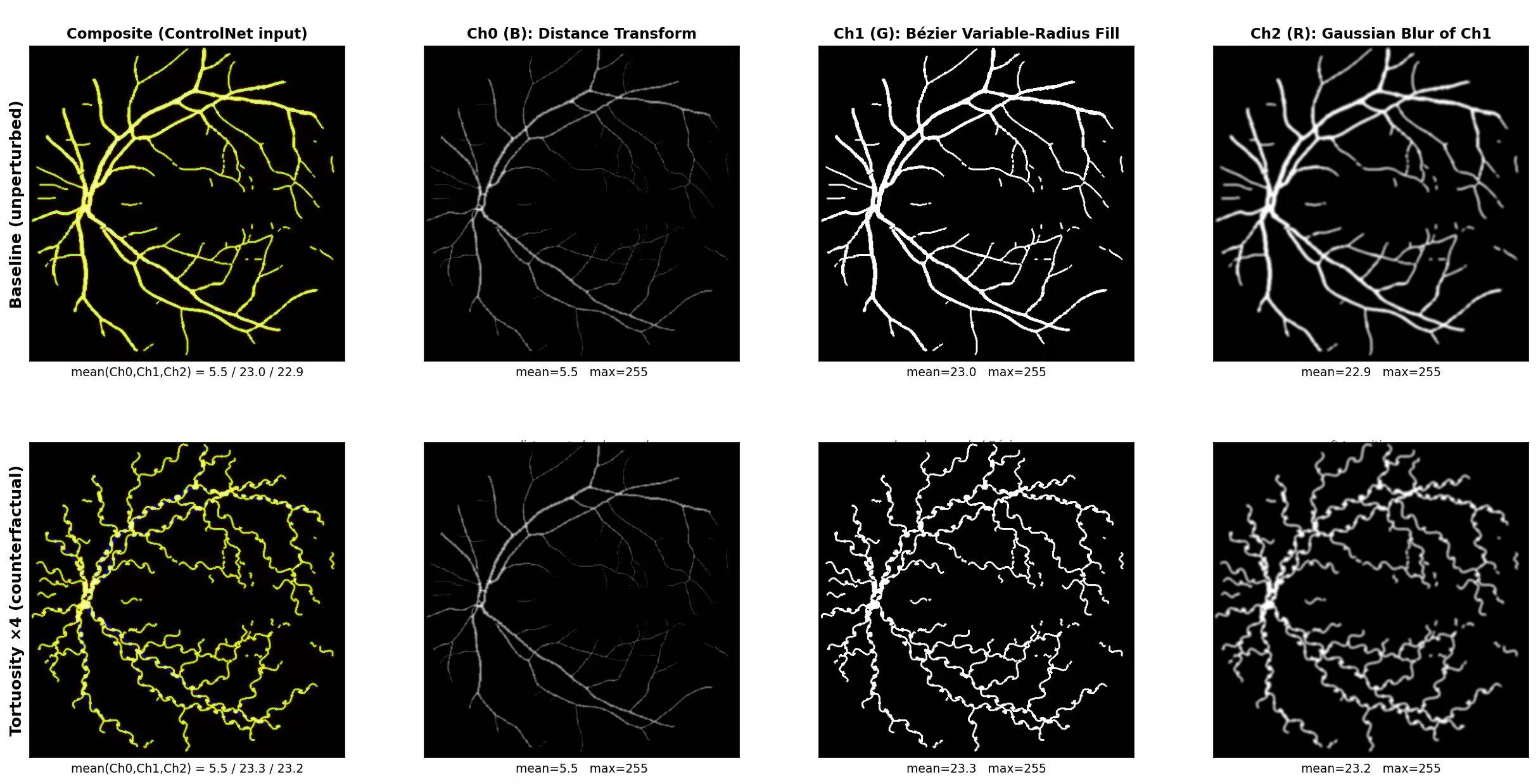}
  \caption{Three-channel B\'ezier hint decomposition on a representative grade-0 fundus. Top row: baseline (unperturbed). Bottom row: \code{tortuosity_4x} counterfactual. Channel~0 (distance-transform radius field) has identical mean ($5.5$ on the $0$--$255$ scale) across both rows, illustrating the construction-level invariance; Channels~1 and~2 visibly distort with the tortuosity perturbation.}
  \label{fig:hint-invariance}
\end{figure}

\paragraph{Channel 0 invariance}
A central design property of the hint is that Channel~0 is invariant under tortuosity and arc-drop by construction: these perturbations modify only the B\'ezier control points, not the distance-transform map from which Channel~0 is computed. Empirically, the mean of Channel~0 differs between baseline and tortuosity-amplified hints by less than $0.022$ on the $[0, 1]$ scale, so the reported tortuosity and arc-drop effects cannot originate from hidden caliber drift. Figure~\ref{fig:hint-invariance} illustrates this side-by-side on a representative grade-0 fundus.

\paragraph{Anatomical coherence}
The tortuosity perturbation is applied per B\'ezier segment and preserves both endpoints $P_0$ and $P_3$ by construction, so vascular connectivity at branch nodes is invariant across all perturbation strengths. The displacement coefficient $\gamma = 0.15$ scales per-segment displacement with the segment chord length, so at \code{tortuosity_4x} per-segment displacement remains within the range of clinically observed retinal vessel tortuosity. Channel~1 (the B\'ezier variable-radius render) draws radii from the invariant Channel~0 to dilate each sampled point of the perturbed skeleton, and the union over all rendered segments absorbs any minor inter-segment overlap into a single radius value rather than producing topologically invalid pixels. The Gaussian-smoothed Channel~2 (kernel $7 \times 7$, $\sigma = 2.0$) further regularizes high-frequency curvature artifacts before the hint reaches the ControlNet. The hint therefore stays in the photorealistic, anatomically coherent range that the LoRA-adapted ControlNet was trained on.

\subsection{Training Details}
\label{app:training}

\paragraph{LoRA finetuning}
We start from the Stable Diffusion 2.1 base and finetune the UNet with a LoRA adapter of rank~32, trained for 36{,}000 steps with AdamW (learning rate $1 \times 10^{-4}$), batch size~8, image size $512 \times 512$. The adapter is applied to the attention and cross-attention projection weights of all transformer blocks. We enable \code{noise_offset = 0.1}: a per-channel spatially-constant offset added to the Gaussian noise during training. In our fundus finetuning runs, omitting this offset made the UNet more prone to gray or bluish color drift rather than the expected red-orange-brown fundus tone. Co-training the LoRA \code{noise_offset} with the ControlNet target in the same batch is required to maintain this calibration; an old LoRA paired with a freshly
trained ControlNet reintroduces the drift. The same coupling rule applies under cross-disease retraining.  

\paragraph{VesselControlNet}
Following the original ControlNet paradigm~\citep{zhang2023adding}, we duplicate the SD~2.1 UNet encoder, attach zero-convolution residuals (convolution layers initialized to zero so the conditioning signal starts from the exact pretrained UNet and gradually leaks in residuals during training), and inject these residuals back into the corresponding UNet decoder blocks.  The checkpoint takes the parametric B\'ezier hint described above as its conditioning input and is used by the counterfactual intervention.

\paragraph{Prompt pairing during training (DR case)}
For each training sample we draw a text prompt uniformly at random from three paraphrases of the sample's class:
\begin{itemize}                              \item Grade~0 (normal):                 
    \begin{itemize}                            \item \code{"a normal fundus photograph without diabetic retinopathy"}             \item \code{"a fundus image of a healthy retina with no lesions"}
    \item \code{"a retinal photograph showing no signs of diabetic retinopathy"}         \end{itemize}                             
\item Grades 1--4 (DR):                        \begin{itemize}
    \item \code{"a fundus photograph with diabetic retinopathy showing microaneurysms and hemorrhages"}           \item \code{"a fundus image showing signs of diabetic retinopathy with retinal lesions"}
    \item \code{"a retinal photograph with diabetic retinopathy featuring exudates and hemorrhages"}                          \end{itemize}                             
\end{itemize}                               
At inference we reproduce this distribution by sampling a paraphrase uniformly at random per synthesized image.

\paragraph{BiomedCLIP encoder for discriminative features}
The discriminative-evaluation features come from a BiomedCLIP encoder~\citep{zhang2023biomedclip} (PubMedBERT $+$ ViT-B/16) fine-tuned on disease labels of the target cohort. Under cross-disease transfer this module is re-fine-tuned with target-disease labels.

\subsection{Perturbation-Axis Selection}
\label{app:axes}

The four selection grounds in \S\ref{sec:perturb-axes} each constrain the candidate axis set; their intersection picks tortuosity, arc length, and radius from the 20-dimensional B\'ezier feature space. We expand each ground in turn below.

\emph{Observational analysis.}
The three perturbed axes all sit within the SHAP top-5 features of the leak-free DR cohort (Appendix~\ref{app:track1-full}, Table~\ref{tab:track1-full}): \#1 \code{arc_length}, \#3 \code{radius}, \#5 \code{tortuosity}. The two skipped top-5 features (\#2 \code{branching_density} and \#4 \code{mean_chord_length}) are both ruled out on the atomic-perturbability ground below.

\emph{Atomic perturbability.}
A counterfactual intervention requires a local operation that moves only the target axis. \code{tortuosity} is atomically perturbable via $P_1, P_2$ displacement (caliber and segment connectivity preserved); \code{arc_drop} is atomically perturbable by deleting fitted segments; \code{radius_x} is atomically perturbable by global \code{dist_map} scaling. These three axes span a natural three-dimensional geometric basis (shape $\perp$ completeness $\perp$ thickness); every other feature in the 20-dimensional set is a functional derivative of this basis. \code{branching_density} cannot be modified without also dropping segments (any branch removal is by definition also an \code{arc_drop}), and similar collinearities rule out \code{mean_chord_length}, \code{mean_curvature}, \code{n_segments}, and \code{min/max_radius} as stand-alone axes.

\emph{Cross-disease literature correspondence.}
For tortuosity, \citet{forster2021retinal} (T2D, $n = 718$, venular tortuosity to incident DR, OR = 1.51) provides direct DR evidence; cross-disease consistency comes from fundus-based AD retinal-vascular biomarker work~\citep{frost2013retinal} and broader retinal vascular evidence in neurodegeneration and stroke~\citep{lemmens2020systematic}. \citet{sandoval2021retinal} further links arteriolar tortuosity and fractal dimension to long-term cardiovascular outcomes in type 2 diabetes. For \code{arc_drop} as a fractal-dimension proxy, \citet{forster2021retinal} (venular FD OR = 0.75), \citet{sandoval2021retinal} (arteriolar FD HR = 0.73), \citet{doubal2010fractal} (decreased FD to lacunar stroke), and the systematic review by \citet{lemmens2020systematic} all support reduced retinal vascular complexity as a neurovascular disease marker. For \code{radius_x}, \citet{klein2004relation}, \citet{cheung2008retinal}, \citet{velayutham2020extended}, and \citet{xu2026quantitative} report early dilation in DR; \citet{wong2004retinal} and \citet{wong2001retinal} report arteriolar narrowing for hypertension and incident stroke. The opposite directions across diseases motivate why the intervention effect must be scored by a disease-specific classifier.

\emph{Dose-response quantifiability.}
Each axis admits three graded perturbation strengths (\code{tort 1x / 2x / 4x}; \code{arc_drop} doses $10 / 20 / 30\%$; \code{radius_x 0.85 / 0.70 / 0.55}), enabling explicit dose-response analyses.

\subsection{Downstream Classifier Hyperparameters}
\label{app:classifier}

 Three backbones are trained with identical hyperparameters: 15 epochs, batch~64, AdamW, learning rate $3 \times 10^{-4}$, cosine schedule, seed~42, $384 \times 384$ ImageNet-normalized input. EfficientNet-B2~@~384 (test AUC 0.9612, sens 0.8858) is the primary scorer; ResNet-50 (AUC 0.9545, sens 0.8320) is the cross-backbone sensitivity check; ViT-B/16 (AUC 0.8295, sens 0.5553) underperforms at the current training scale, and we treat its counterfactual-intervention output as diagnostic rather than confirmatory. The stroke and AD EfficientNet-B2 heads are retrained from scratch on their respective cohorts with the same hyperparameter family; numbers are in \S\ref{sec:cross-disease}.

\subsection{Inference Hyperparameters and Filter Thresholds}
\label{app:inference}

A single fixed set of sampling hyperparameters is used across all counterfactual intervention experiments to ensure within-start paired $\Delta$ comparability; the guidance-weight ablation (\S\ref{sec:cf-guidance}) sweeps CFG weight $w$ as the sole exception. Table~\ref{tab:inference-hparams} summarizes the values.

\begin{table}[!htbp]
  \centering
  \caption{Inference hyperparameters used uniformly across the counterfactual intervention experiments.}
  \label{tab:inference-hparams}
  \resizebox{\linewidth}{!}{%
  \begin{tabular}{lll}
    \toprule
    Parameter & Value & Role \\
    \midrule
    DDIM steps & 50 & Sampling schedule \\
    CFG \code{guidance_scale} & 7.5 & Text-conditioning strength \\
    ControlNet \code{cn_scale} & 1.0 & Residual strength \\
    \code{mid_scale} & 0.5 & Mid-block residual strength \\
    \code{latent_offset} & 0.3 & Initial latent bias toward real-fundus latent mean \\
    LoRA weights & v2 (\code{noise_offset = 0.1}) & Color-calibrated adapter \\
    \bottomrule
  \end{tabular}}
\end{table}

\paragraph{Visual-fidelity filter thresholds}
The visual-fidelity filter referenced in \S\ref{sec:cf-design} discards baseline ControlNet outputs that fail to render as a valid fundus. A start is retained only if its baseline generation satisfies all three conditions: (i) mean pixel intensity in $[50, 170]$ on the $0$--$255$ scale, rejecting near-black collapses (mean $< 50$) and saturated bright-white collapses (mean $> 170$); (ii) pixel standard deviation $> 25$, rejecting monochromatic collapses where the entire image converges to a single tone; and (iii) red-to-green channel ratio $> 1.3$, enforcing the red-orange-brown chromaticity of healthy fundi and rejecting gray or bluish drift. On the $n = 500$ DR cohort, $350$ of $500$ starts pass the visual-fidelity filter; further restricting to baseline DR probability $< 0.3$ yields $n = 190$ under DR-prompt and $n = 97$ under Uncond.

\subsection{Dataset Deduplication}                         \label{app:deduplication}                              We enforce strict subject-level deduplication via two operations:
  \begin{itemize}                                           \item \textbf{Cross-split anti-leakage.} From the train split, we remove $1{,}032$ rows whose \code{base_id} appears in the test split, ensuring that no patient crosses the train/test boundary. \item \textbf{In-split augmentation collapse.} The
 manual augmentation of the redistribution produces multiple synthetic copies that share the
 same \code{base_id} per patient. Within val and test, we collapse
 these into a single representative per patient ($958$ rows removed from val and $964$ from test), so that evaluation metrics reflect patient-level rather than image-level performance.
  \end{itemize}                                                                                     
  Together the two operations remove $1{,}032 + 958 + 964 = 2{,}954$ rows from the original $143{,}669$-image redistribution, yielding the final splits $114{,}209 / 13{,}269 / 13{,}237$ (train / val / test) reported in \S\ref{sec:data-setup}.


\section{Extended Empirical Analyses}

\subsection{Generator Visual Bias and the Counterfactual Probe}
\label{app:visual-bias}

A natural concern is that the generator's residual visual bias contaminates the counterfactual-intervention measurement. Four empirical controls address this concern.

\paragraph{Pixel-level noise sensitivity}
The classifier does not respond to simple color noise: \code{pixdrop_30} yields only $0.004$ (DR-prompt) / $0.054$ (Uncond), $16$ to $100$ times weaker than \code{tortuosity_4x}. The low response to mildly-artifacted baselines indicates that the classifier discriminates on pathologically-meaningful structure, not pixel-level noise.

\paragraph{Style versus content}
Counterfactual purity rests on the within-start paired $\Delta$ design rather than on the generator's absolute content fidelity. For each starting fundus, the baseline and perturbed generations share the same start, the same generator, and the same text prompt; they differ only in the controlled parametric perturbation of the B\'ezier hint. Any systematic style bias (tone, texture, vessel-geometry idiosyncrasies of the generator) therefore enters baseline and perturbed identically and cancels as a common term in $\Delta$. Whether the generator produces photorealistically correct DR content (microaneurysms, hemorrhages, exudates) is irrelevant to the comparability of the paired difference: only the \emph{differential} response to the parametric perturbation enters $\Delta$. The matched pixel-drop control (Appendix~\ref{app:track4-full-table}) further confirms that $\Delta$ tracks the geometric intervention rather than generator-side artifacts. 

\paragraph{Pixel-drop contrast}
The three-way contrast among Baseline ($0.097$), \code{tortuosity_4x} ($0.437$), and \code{pixdrop_30} ($0.004$) isolates the causal driver as topological geometry rather than pixel count, low-level pixel noise, or tone.

\subsection{Observational Analysis: Full 20-Feature Table}
\label{app:track1-full}

Table~\ref{tab:track1-full} reports the complete 20-dimensional B\'ezier observational analysis on the DR cohort ($n = 26{,}506$).

\begin{table}[!htbp]
  \centering
  \footnotesize
  \caption{Full 20-feature B\'ezier observational analysis on the DR cohort, ordered by SHAP importance. Metrics defined in \S\ref{sec:eval-protocols}; \emph{SHAP non-linear} flags features failing the Spearman--SHAP sign-disagreement criterion. Significance: $***$ $p < 0.001$, $*$ $p < 0.05$.}
  \label{tab:track1-full}
  \resizebox{\linewidth}{!}{%
  \begin{tabular}{clccccll}
    \toprule
    Rank & Feature & Spearman $\rho$ & feat-SHAP corr & OR/SD & Q5/Q1 OR & SHAP abs & Direction \\
    \midrule
    1  & \code{total_arc_length}      & $-0.358$***   & $-0.94$ & 0.55 & 0.20  & 0.456 & $\downarrow$ DR \\
    2  & \code{branching_density}     & $+0.014$*     & $-0.55$ & 1.21 & 1.09  & 0.364 & SHAP non-linear \\
    3  & \code{mean_radius}           & $-0.084$***   & $-0.90$ & 0.85 & 0.68  & 0.285 & $\downarrow$ DR \\
    4  & \code{mean_chord_length}     & $-0.079$***   & $-0.84$ & 0.80 & 0.67  & 0.231 & $\downarrow$ DR \\
    5  & \code{mean_tortuosity}       & $+0.086$***   & $+0.17$ & 1.00 & 1.60  & 0.160 & $\uparrow$ DR \\
    6  & \code{mean_segment_length}   & $-0.077$***   & $-0.81$ & 0.80 & 0.67  & 0.141 & $\downarrow$ DR \\
    7  & \code{radius_cv}             & $+0.007$ n.s. & $+0.74$ & 0.94 & 1.29  & 0.135 & $\uparrow$ DR \\
    8  & \code{std_radius}            & $-0.028$***   & $+0.30$ & 0.91 & 1.04  & 0.102 & SHAP non-linear \\
    9  & \code{std_chord_length}      & $+0.087$***   & $-0.49$ & 0.99 & 1.53  & 0.071 & SHAP non-linear \\
    10 & \code{std_tortuosity}        & $+0.078$***   & $-0.51$ & 1.17 & 1.55  & 0.063 & SHAP non-linear \\
    11 & \code{max_radius}            & $-0.120$***   & $-0.68$ & 0.84 & 0.62  & 0.062 & $\downarrow$ DR \\
    12 & \code{thick_vessel_ratio}    & $-0.033$***   & $+0.20$ & 1.06 & 0.96  & 0.050 & SHAP non-linear \\
    13 & \code{mean_curvature}        & $+0.091$***   & $-0.64$ & 1.22 & 1.65  & 0.049 & SHAP non-linear \\
    14 & \code{std_segment_length}    & $+0.087$***   & $+0.35$ & 0.99 & 1.55  & 0.046 & $\uparrow$ DR \\
    15 & \code{n_segments}            & $-0.338$***   & $-0.57$ & 0.58 & 0.23  & 0.035 & $\downarrow$ DR \\
    16 & \code{coverage_ratio}        & $-0.315$***   & $+0.39$ & 0.59 & 0.25  & 0.034 & SHAP non-linear \\
    17 & \code{std_curvature}         & $+0.102$***   & $-0.18$ & 1.21 & 1.72  & 0.028 & SHAP non-linear \\
    18 & \code{max_tortuosity}        & $-0.026$***   & $+0.37$ & 1.08 & 0.98  & 0.026 & SHAP non-linear \\
    19 & \code{max_curvature}         & $-0.043$***   & $+0.29$ & 1.04 & 1.01  & 0.025 & SHAP non-linear \\
    20 & \code{min_radius}            & $+0.123$***   & $+0.83$ & 1.17 & $\mathbf{2.55}$ & 0.005 & $\uparrow$ DR \\
    \bottomrule
  \end{tabular}}
\end{table}


\section{Full Counterfactual Intervention Causal Table}
\label{app:track4-full-table}

\begin{figure}[!htbp]
  \centering
  \includegraphics[width=0.5\linewidth]{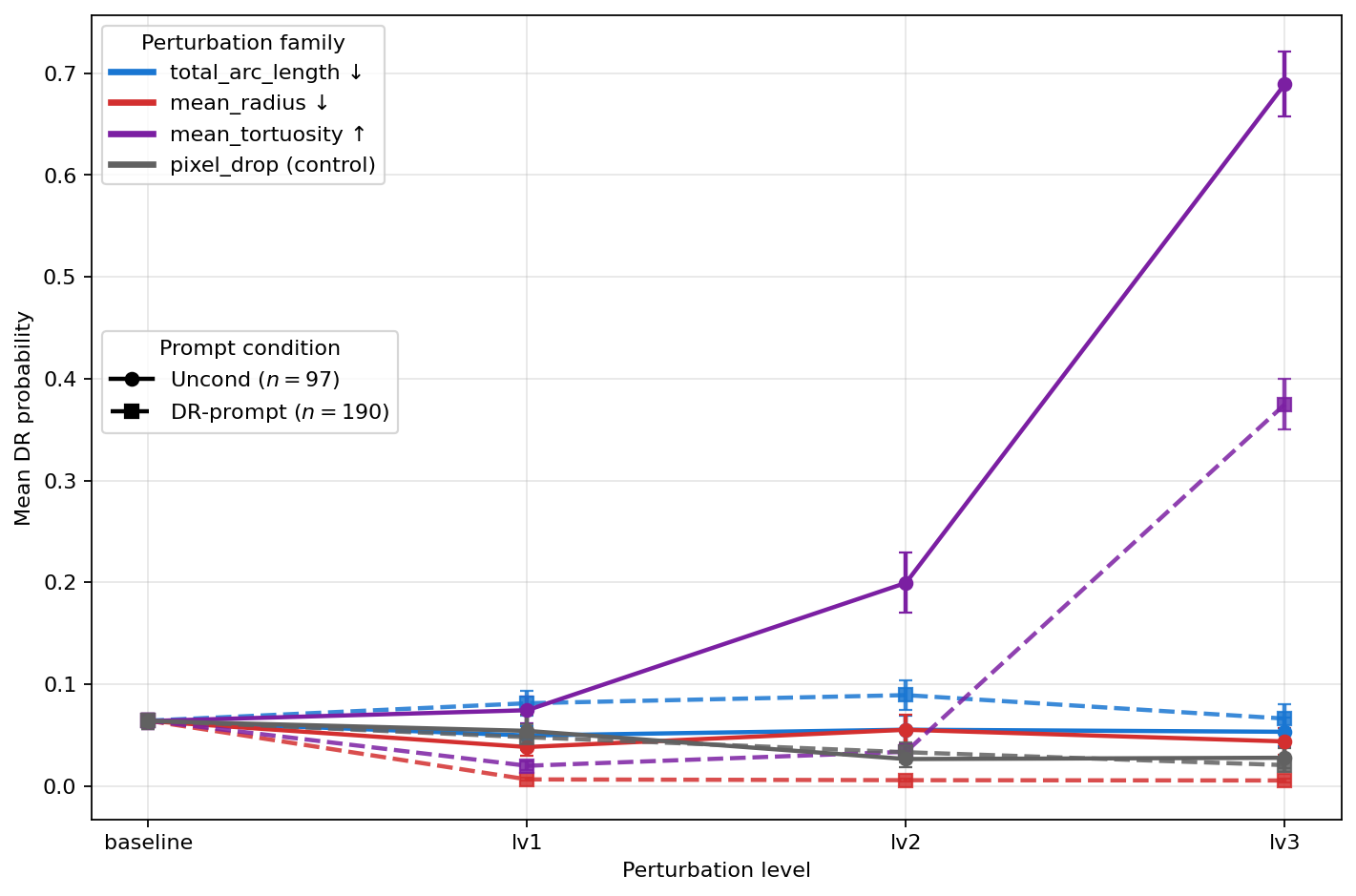}                  \caption{Causal-response curves: mean DR probability across all 13 perturbation configurations under DR-prompt and Uncond. Tortuosity is monotonic; \code{pixdrop} saturates near zero. Numerical values for all configurations are in Table~\ref{tab:track4-full}.}                       \label{fig:causal-curves}                                               \end{figure}

Figure~\ref{fig:causal-curves} visualizes mean DR probability across all 13 perturbation configurations under DR-prompt and Uncond; Table~\ref{tab:track4-full} reports the corresponding paired $\Delta$ values. Five patterns recur. (i) The strongest tortuosity dose yields the largest effect under both prompt conditions, with the Uncond response $2.0\times$ stronger than the DR-prompt response, visible as the steeply rising tortuosity curve in Figure~\ref{fig:causal-curves}. (ii) The mid-dose tortuosity reveals an Uncond-only effect that is hidden under the DR-prompt. (iii) The arc-drop family is near-null at all three doses, indicating that segment removal alone does not trigger the classifier when radius and topology are preserved. (iv) The radius family yields a small consistent negative $\Delta$ in agreement with the mainstream DR direction under both conditions. (v) The pixel-drop control yields no statistically significant effect at any of the three doses, with the strongest pixel-drop $\sim\!18\times{/}25\times$ weaker than \code{tortuosity_4x}. Figure~\ref{fig:causal-curves} shows the \code{pixdrop} trace saturating near zero throughout.

\section{Cross-Disease Summary}
\label{app:cross-disease-summary}

Table~\ref{tab:three-disease} compares the framework's three diseases (DR primary case, ischemic stroke, Alzheimer's disease) on cohort scale, observational top signals, discriminative AUC across feature sets, and the strongest counterfactual effect. Effect-size attenuation across DR $>$ stroke $>$ AD tracks both cohort scale and discriminative-baseline strength, as expected; the AD cohort additionally inverts the discriminative ranking (B\'ezier $>$ CLIP), supporting the interpretation that the AD signal is captured by vessel geometry.

\begin{table}[!htbp]
  \centering
  \footnotesize
  \caption{Cross-disease summary across the three diseases. Image-level AUC refers to a disease-specific EfficientNet-B2 classifier trained end-to-end. The counterfactual row reports $\Delta$ on \code{tortuosity_4x} on the strict reverse-counterfactual subset (baseline disease probability $< 0.3$).}
  \label{tab:three-disease}
  \begin{tabular}{lccc}
    \toprule
     & DR (primary) & Ischemic Stroke & Alzheimer's Disease \\
    \midrule
    \multicolumn{4}{l}{\emph{Cohort and prevalence}} \\
    Eval cohort $n$ & $26{,}506$ & $3{,}409$ & $550$ \\
    Positive prevalence & $\sim\!49\%$ (grades 1--4) & $12\%$ & $7.5\%$ \\
    \midrule
    \multicolumn{4}{l}{\emph{Observational analysis ($p<0.05$ Spearman)}} \\
    Significant features ($/20$) & $19$ & $9$ & $1$ \\
    Top Q5/Q1 OR (feature) & $\mathbf{5.0}$ (\code{min_radius}) & $0.62$ (\code{n_segments}) & $\mathbf{3.84}$ (\code{std_curvature}) \\
    \midrule
    \multicolumn{4}{l}{\emph{Discriminative test AUC}} \\
    B\'ezier (20-d) & $0.706$ & $0.536$ & $\mathbf{0.646}$ \\
    BiomedCLIP (512-d) & $\mathbf{0.836}$ & $0.528$ & $0.454$ \\
    Image-level (EffB2) & $\mathbf{0.961}$ & $0.534$ & $0.624$ \\
    \midrule
    \multicolumn{4}{l}{\emph{Counterfactual $\Delta$ on \code{tortuosity_4x}}} \\
    $\Delta$ (Uncond) & $\mathbf{+0.625}$ & --- & --- \\
    $\Delta$ (DR-/disease-prompt) & $+0.311$ & $+0.337$ & $+0.191$ \\
    Strict-subset $n$ & $97$ / $190$ & $17$ & $30$ \\
    Pixdrop control / Tort.\ ratio & $\sim\!1/18$--$1/25$ & $\sim\!1/200$ & $\sim\!1/1$ \\
    \bottomrule
  \end{tabular}
\end{table}

\subsection{Stroke Full Tables}
\label{app:cross-disease-full}

\paragraph{Cohort-design limitation and its effect on discriminative AUC}
The stroke cohort admits past, concurrent, and post-imaging stroke events under a single ICD-coded ``stroke history'' label, so the discrimination task is closer to ``has this participant ever had or will they have a stroke'' than to a clean incident-prediction task. A single fundus image carries far less signal for past stroke events because the retina need not retain a permanent imprint of a remote infarction; most past-stroke participants are therefore visually indistinguishable from unaffected controls. This is the structural reason behind the near-chance discriminative numbers: XGBoost yields B\'ezier (20-d) test AUC $0.536$, BiomedCLIP frozen (512-d) $0.528$; a single-view stroke CNN reaches AUC $0.534$, sensitivity $0.205$, specificity $0.845$.

\paragraph{Comparison with the most relevant UK Biobank stroke benchmark}
\citet{yusufu2025retinal} report an adjusted AUC of $0.752$ on UK Biobank ($n = 45{,}161$, $749$ incident strokes) using $118$ retinal vascular parameters (calibre, density, tortuosity, branching angle, complexity) extracted by the Retina-based Microvascular Health Assessment System (RMHAS) together with traditional risk factors, vs.\ $0.738$ from risk factors alone. Their cohort excludes prevalent stroke at baseline, restricting the task to incident-only prediction, and the analysis is observational (Cox regression with restricted cubic spline). The authors themselves note that ``this study's observational nature limits the establishment of causality between retinal vascular parameters and stroke risk'', which is precisely the gap our retinal-geometry-counterfactual framework targets. The AUC difference therefore reflects the incident-only vs.\ mixed-temporal design contrast, not a representation failure, and we treat our discriminative AUC as a task-design floor.

\paragraph{Counterfactual results align with prior observations}
Even at this discriminative floor, parameter-level intervention on the strongest tortuosity dose (\code{tortuosity_4x}) reproduces a clean monotonic dose-response (Table~\ref{tab:stroke-track4-full}); Spearman regression on the 20-dimensional B\'ezier feature set yields nine features at $p < 0.05$ (Table~\ref{tab:stroke-track1-full}), all in the down-stroke direction, six of which align directionally with caliber-narrowing~\citep{wong2001retinal} and fractal-dimension reduction~\citep{doubal2010fractal}.

\begin{table}[!htbp]
  \centering
  \footnotesize
  \caption{Stroke observational analysis: 20-dimensional B\'ezier feature association on the $n = 3{,}409$ evaluation cohort. Significance flags use the convention $* p<0.05$, $** p<0.01$, $*** p<0.001$. The nine features in the upper block are significant at $p < 0.05$; the lower block is non-significant.}
  \label{tab:stroke-track1-full}
  \begin{tabular}{lccc}
    \toprule
    Feature & Spearman $\rho$ & OR/SD (95\% CI) & Q5/Q1 OR \\
    \midrule
    \code{n_segments} & $-0.065$~*** & $0.819$ ($0.74$, $0.91$) & $0.62$ \\
    \code{total_arc_length} & $-0.062$~*** & $0.831$ ($0.75$, $0.92$) & $0.66$ \\
    \code{coverage_ratio} & $-0.058$~*** & $0.839$ ($0.76$, $0.93$) & $0.68$ \\
    \code{max_radius} & $-0.050$~** & $0.859$ ($0.78$, $0.95$) & $0.62$ \\
    \code{std_radius} & $-0.040$~* & $0.885$ ($0.80$, $0.98$) & $0.79$ \\
    \code{radius_cv} & $-0.039$~* & $0.892$ ($0.81$, $0.98$) & $0.71$ \\
    \code{mean_radius} & $-0.038$~* & $0.885$ ($0.81$, $0.97$) & $0.84$ \\
    \code{max_curvature} & $-0.038$~* & $0.883$ ($0.78$, $1.00$) & $0.97$ \\
    \code{thick_vessel_ratio} & $-0.035$~* & $0.904$ ($0.81$, $1.01$) & $0.82$ \\
    \midrule
    \code{std_curvature} & $-0.033$ n.s.\ & $0.888$ ($0.80$, $0.98$) & $0.79$ \\
    \code{mean_curvature} & $-0.033$ n.s.\ & $0.895$ ($0.79$, $1.02$) & $0.72$ \\
    \code{mean_segment_length} & $+0.032$ n.s.\ & $0.942$ ($0.86$, $1.04$) & $1.28$ \\
    \code{mean_chord_length} & $+0.031$ n.s.\ & $0.942$ ($0.86$, $1.04$) & $1.23$ \\
    \code{min_radius} & $-0.023$ n.s.\ & $0.909$ ($0.83$, $1.00$) & $0.94$ \\
    \code{max_tortuosity} & $-0.017$ n.s.\ & $0.938$ ($0.84$, $1.05$) & $0.93$ \\
    \code{std_chord_length} & $-0.009$ n.s.\ & $0.918$ ($0.84$, $1.01$) & $0.90$ \\
    \code{std_segment_length} & $-0.008$ n.s.\ & $0.919$ ($0.84$, $1.01$) & $0.92$ \\
    \code{std_tortuosity} & $-0.008$ n.s.\ & $0.979$ ($0.87$, $1.10$) & $1.07$ \\
    \code{mean_tortuosity} & $+0.002$ n.s.\ & $0.917$ ($0.84$, $1.00$) & $0.92$ \\
    \code{branching_density} & $-0.001$ n.s.\ & $1.007$ ($0.91$, $1.12$) & $0.95$ \\
    \bottomrule
  \end{tabular}
\end{table}

\begin{table}[!htbp]
  \centering
  \footnotesize
  \caption{Stroke counterfactual intervention: 13-configuration causal table on the $n = 17$ strict reverse-counterfactual subset (baseline stroke probability~$< 0.3$). Mean $\pm$ SEM and $95\%$ CI on stroke probability under the EfficientNet-B2 stroke head. Configurations sorted by mean post-intervention probability.}
  \label{tab:stroke-track4-full}
  \begin{tabular}{lcccc}
    \toprule
    Configuration & Mean & SEM & 95\% CI lower & 95\% CI upper \\
    \midrule
    baseline & $0.0009$ & $0.0003$ & $0.0002$ & $0.0016$ \\
    \code{pixdrop_30} & $0.0017$ & $0.0014$ & $-0.0013$ & $0.0047$ \\
    \code{tortuosity_1x} & $0.0019$ & $0.0008$ & $0.0002$ & $0.0036$ \\
    \code{arc_drop_10} & $0.0027$ & $0.0016$ & $-0.0007$ & $0.0061$ \\
    \code{radius_x0.85} & $0.0028$ & $0.0016$ & $-0.0006$ & $0.0062$ \\
    \code{arc_drop_20} & $0.0030$ & $0.0015$ & $-0.0002$ & $0.0062$ \\
    \code{pixdrop_10} & $0.0053$ & $0.0035$ & $-0.0021$ & $0.0127$ \\
    \code{pixdrop_20} & $0.0065$ & $0.0052$ & $-0.0045$ & $0.0175$ \\
    \code{radius_x0.70} & $0.0115$ & $0.0096$ & $-0.0089$ & $0.0319$ \\
    \code{radius_x0.55} & $0.0163$ & $0.0146$ & $-0.0146$ & $0.0472$ \\
    \code{tortuosity_2x} & $0.0178$ & $0.0087$ & $-0.0006$ & $0.0362$ \\
    \code{arc_drop_30} & $0.0278$ & $0.0173$ & $-0.0089$ & $0.0645$ \\
    \code{tortuosity_4x} & $\mathbf{0.3383}$ & $0.0716$ & $0.1865$ & $0.4901$ \\
    \bottomrule
  \end{tabular}
\end{table}

\subsection{Alzheimer's Disease Full Tables}
\label{app:ad-full}

\paragraph{Cohort, discriminative-ranking inversion, and AD classifier}
The Alzheimer's disease (AD) cohort comprises $550$ unique participants with patient-level deduplication (eval split), of whom $41$ ($7.5\%$) carry an AD diagnosis label; training and val/test splits follow the protocol used for DR and stroke. XGBoost on the evaluation cohort yields B\'ezier (20-d) test AUC $0.646$, BiomedCLIP frozen (512-d) $0.454$. The B\'ezier $>$ CLIP ordering inverts that of DR, supporting that the AD signal is captured by parametric vessel geometry rather than by the lesion-pathological visual content. The single-view AD CNN used as the counterfactual scorer (EfficientNet-B2, weighted cross-entropy, $30$ epochs) achieves test AUC $0.624$, sensitivity $0.50$, specificity $0.677$.

\paragraph{Observational anchor}
Spearman regression on the 20-dimensional B\'ezier feature set yields \code{std_curvature} significant at $p < 0.05$ ($\rho = +0.086$, OR/SD $= 1.41$ [$1.04$, $1.92$]) with the largest Q5/Q1 odds ratio across the three diseases ($3.84$, $95\%$ CI $[1.30, 11.37]$), consistent with~\citet{frost2013retinal}, who report fundus-based retinal vascular biomarker differences in Alzheimer's disease.

\paragraph{Counterfactual intervention}
On the $n = 30$ strict reverse-counterfactual subset (baseline AD probability $<0.3$), parameter-level intervention on \code{tortuosity_4x} produces the highest mean response within the tortuosity family ($0.299$ vs.\ baseline $0.109$, $\Delta = +0.19$), reproducing the positive tortuosity direction observed on DR and stroke. The 13-configuration causal table is in Table~\ref{tab:ad-track4-full}; the specificity caveat tied to the AD classifier is discussed in \S\ref{sec:cross-disease}.

\begin{table}[!htbp]
  \centering
  \footnotesize
  \caption{AD counterfactual intervention: 13-configuration causal table on the $n = 30$ strict reverse-counterfactual subset (baseline AD probability~$< 0.3$). Mean and SEM on AD probability under the AD-retrained EfficientNet-B2 head; two-sided $95\%$ CI computed using Student's $t$ distribution with $df = 29$ ($t_{0.025} = 2.045$). Configurations sorted by mean post-intervention probability.}
  \label{tab:ad-track4-full}
  \begin{tabular}{lcccc}
    \toprule
    Configuration & Mean & SEM & 95\% CI lower & 95\% CI upper \\
    \midrule
    baseline & $0.109$ & $0.021$ & $0.066$ & $0.152$ \\
    \code{pixdrop_30} & $0.169$ & $0.046$ & $0.075$ & $0.263$ \\
    \code{arc_drop_10} & $0.182$ & $0.034$ & $0.112$ & $0.252$ \\
    \code{radius_x0.85} & $0.202$ & $0.044$ & $0.112$ & $0.292$ \\
    \code{tortuosity_2x} & $0.202$ & $0.040$ & $0.120$ & $0.284$ \\
    \code{pixdrop_20} & $0.205$ & $0.050$ & $0.103$ & $0.307$ \\
    \code{tortuosity_1x} & $0.207$ & $0.044$ & $0.117$ & $0.297$ \\
    \code{radius_x0.70} & $0.252$ & $0.055$ & $0.140$ & $0.364$ \\
    \code{radius_x0.55} & $0.272$ & $0.055$ & $0.160$ & $0.384$ \\
    \code{tortuosity_4x} & $\mathbf{0.299}$ & $0.065$ & $0.166$ & $0.432$ \\
    \code{arc_drop_30} & $0.318$ & $0.059$ & $0.197$ & $0.439$ \\
    \code{pixdrop_10} & $0.323$ & $0.061$ & $0.198$ & $0.448$ \\
    \code{arc_drop_20} & $0.333$ & $0.054$ & $0.223$ & $0.443$ \\
    \bottomrule
  \end{tabular}
\end{table}

\section{Counterfactual Examples and Grad-CAM Saliency}
\label{app:gradcam}

\begin{figure}[!htbp]
  \centering
  \includegraphics[width=0.6\linewidth]{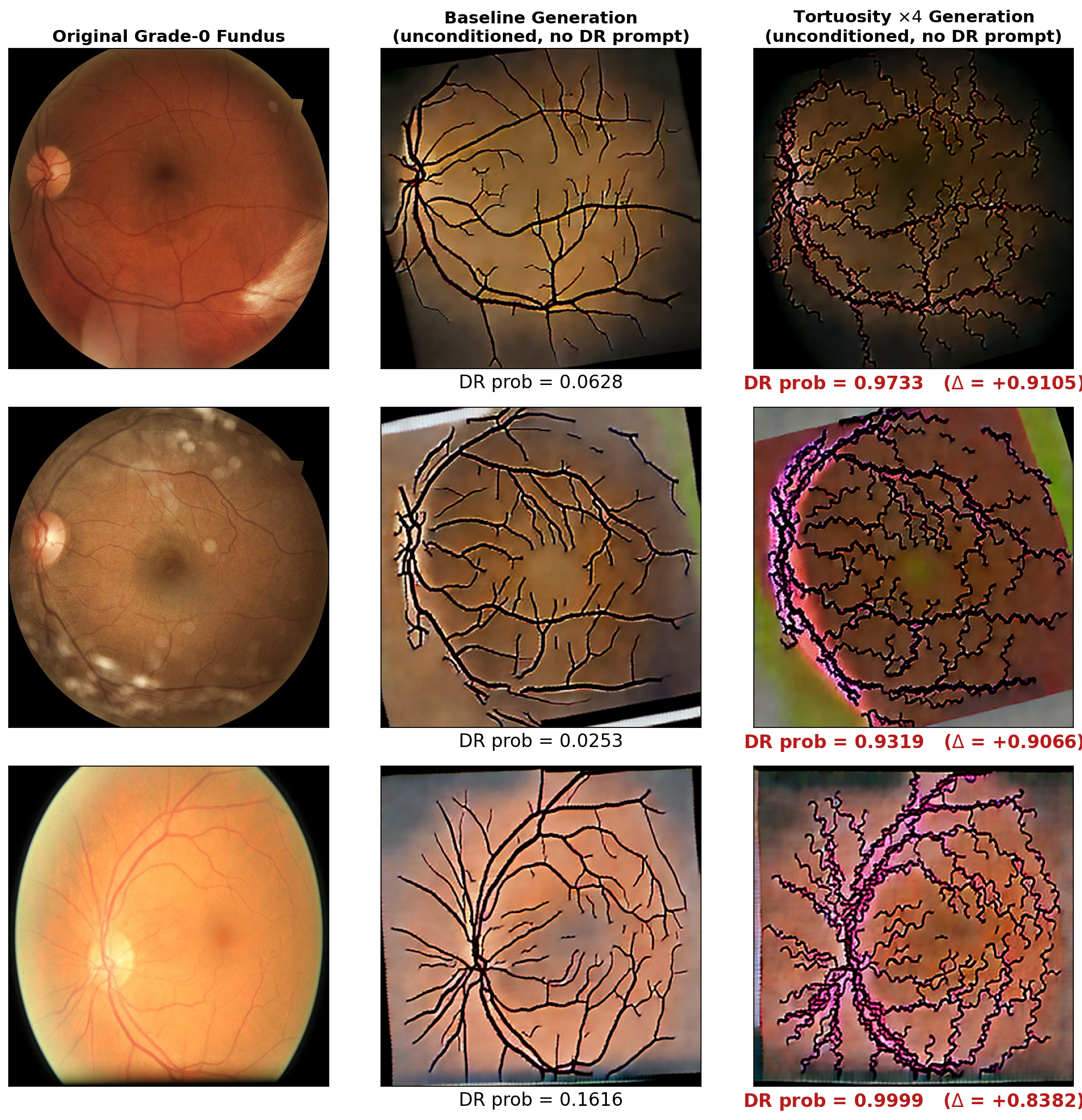}
  \caption{Three counterfactual examples. Each example shifts from baseline DR probability $< 0.17$ to $> 0.93$ under \code{tortuosity_4x} intervention. Each row shows (i) the original grade-0 fundus, (ii) the baseline ControlNet generation, (iii) the \code{tortuosity_4x} generation.}
  \label{fig:champions}
\end{figure}

\begin{figure}[!htbp]
  \centering
  \includegraphics[width=\linewidth]{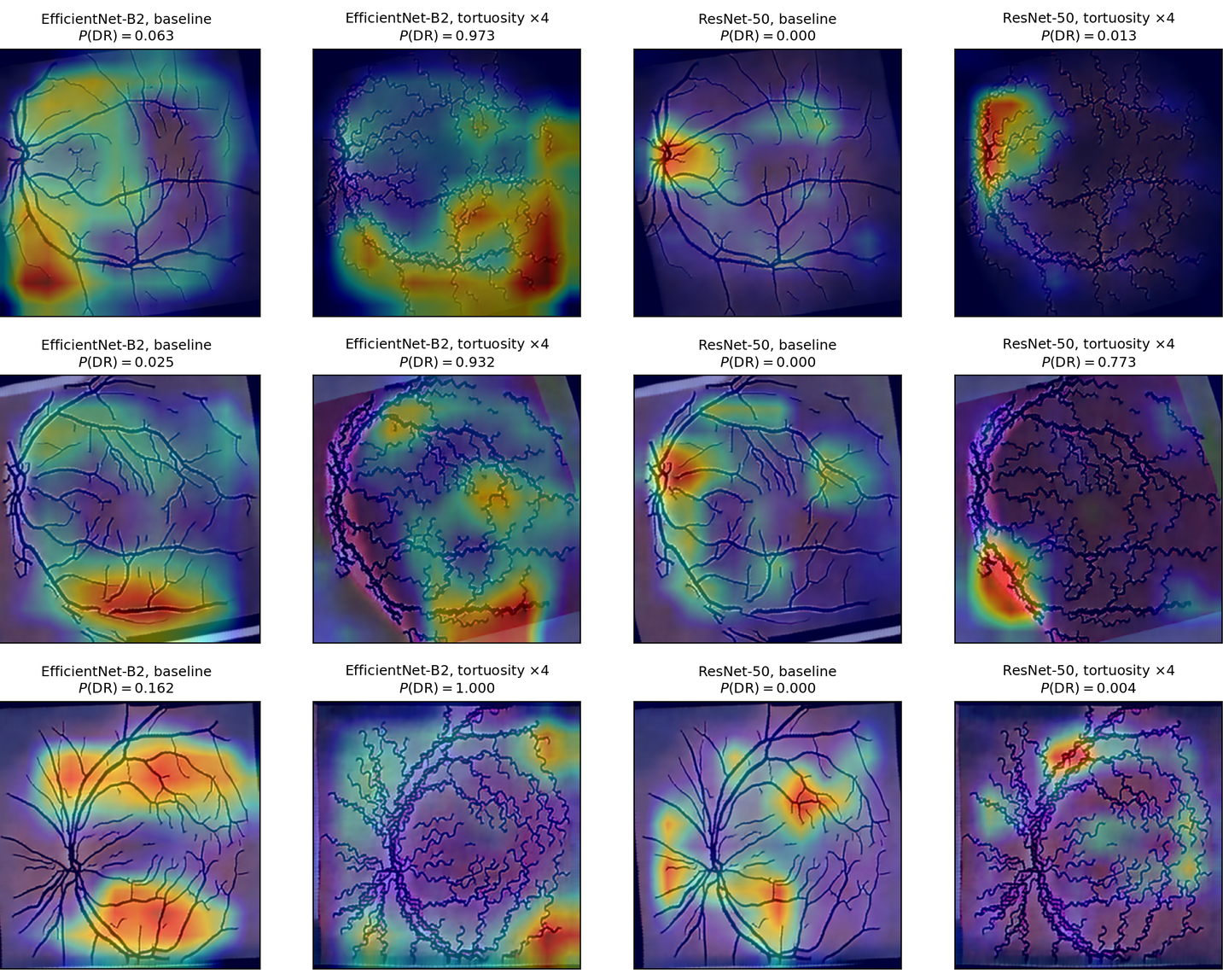}
  \caption{Grad-CAM saliency overlays on the three high-fidelity counterfactual examples (rows) under EfficientNet-B2 and ResNet-50 (columns), comparing baseline against \code{tortuosity_4x}. Heatmaps use the standard Grad-CAM \texttt{jet} colormap with $\alpha = 0.40$ overlay so that vessel topology stays legible underneath. Predicted DR probability $P(\mathrm{DR})$ is annotated per panel. High-activation regions concentrate on the perturbed vessel segments, supporting the topological-geometry-as-causal-driver conclusion.}
  \label{fig:gradcam}
\end{figure}

We first show three high-fidelity counterfactual examples (Figure~\ref{fig:champions}) where the strongest tortuosity dose drives DR probability from baseline $< 0.17$ to $> 0.93$, and then render Grad-CAM saliency maps on the same three examples under both downstream backbones (Figure~\ref{fig:gradcam}).

To confirm that the classifier's elevated DR probability under \code{tortuosity_4x} is driven by the perturbed vessel topology rather than by out-of-distribution generation artifacts, we contrast baseline against \code{tortuosity_4x} (Figure~\ref{fig:gradcam}). Across all three examples and both backbones, the high-activation regions concentrate on the curving vessel segments amplified by the parametric intervention rather than on disc edges, image padding, or generation artifacts. Both the EfficientNet-B2 and ResNet-50 classifiers reproduce the dose-monotonic shift, with predicted DR probability rising from baseline values below $0.17$ to above $0.93$ on the strongest dose. The architecture-agnostic localization of the salient regions on the perturbed vessels, combined with the $16$--$100\times$ pixdrop attenuation reported in \S\ref{app:visual-bias}, supports the conclusion that the classifier responds to topological geometry rather than to OOD-shortcut artifacts.

\section{Conditioning Taxonomy (L0--L3)}
\label{app:l0-l3}

Table~\ref{tab:l0-l3} summarizes the four levels and their representative work. L0 has parametric control but no photorealistic texture. L1 has photorealistic texture but no clean parametric decoupling. L2 is the first bridge; the controllability is parametric on the user-facing upstream while the generator side is unchanged: ControlNet receives a $(512, 512, 3)$ pixel tensor identical in format to a binary mask. L3 is left to future work: a parameter-native ControlNet that takes the B\'ezier control-point sequence directly (e.g.\ as a token sequence attended by the diffusion UNet) would remove the rasterization step and potentially resolve the generator's ControlNet visual bias at its source. The engineering cost is substantial, but a parameter-native ControlNet would benefit all retinal-vessel-mediated diseases addressed by the framework.

\begin{table}[!htbp]
  \centering
  \caption{Four-level taxonomy of conditioning for fundus vessel synthesis. This work occupies the L2 bridge.}
  \label{tab:l0-l3}
  \resizebox{\linewidth}{!}{%
  \begin{tabular}{llll}
    \toprule
    Level & Upstream control & Downstream generator & Representative work \\
    \midrule
    L0 & Classical renderers (ASM, Catmull-Rom) & Non-generative / interactive GUI & Bonaldi 2016~\citep{bonaldi2016automatic}; Castro 2020~\citep{castro2020visual} \\
    L1 & Binary vessel mask (pixel) & Deep generative (GAN / Diffusion) & Guo2024~\citep{guo2024controllable}; Feng 2024~\citep{feng2024diversified}; Go 2024~\citep{go2024generation} \\
    L2~$\star$ & Cubic B\'ezier control points to 3-channel hint & SD~2.1 + LoRA + ControlNet & This work (BTECF) \\
    L3 & B\'ezier parameters as condition embeddings & Re-trained ControlNet / diffusion & Future work \\
    \bottomrule
  \end{tabular}}
\end{table}


\section{DR Pathology Analysis: Three Causal Chains}
\label{app:dr-pathology}

This appendix sets out three histopathological causal chains that together account for the SHAP top-5 directions of Table~\ref{tab:track1-top5} (\S\ref{sec:obs}): \code{total_arc_length} ($\downarrow$ DR), \code{branching_density} (\emph{SHAP non-linear} with weak positive Spearman), \code{mean_radius} ($\downarrow$ DR), \code{mean_chord_length} ($\downarrow$ DR), and \code{mean_tortuosity} ($\uparrow$ DR). The chains share intermediate nodes (pericyte loss, capillary closure, local retinal hypoxia) and jointly account for the five directions as follows: three features (\code{total_arc_length}, \code{mean_radius}, \code{mean_chord_length}) decrease monotonically, one feature (\code{mean_tortuosity}) rises monotonically through a same-direction superposition of an early mechanical arm and a late neovascular arm, and one feature (\code{branching_density}) carries a biphasic signature in which a Chain-1 subtractive arm and a Chain-3 late-stage neovessel-additive arm coexist.

\subsection{Chain 1: Vasoregression and Branch Loss}

The first chain begins at the inner blood--retinal barrier. Hyperglycemia drives selective pericyte loss, which removes mechanical and signaling support from the capillary endothelium and produces endothelial-cell dysfunction. The dysfunctional endothelium is the substrate of two co-occurring lesions: focal microaneurysm formation and segmental capillary closure. Capillary closure produces local retinal ischemia, which drives \emph{vasoregression}, the active deletion of the mature capillary network together with atrophy of the immediately upstream arterioles. The vessels deleted by vasoregression are predominantly the small distal branches that once filled the periphery of the AutoMorph segmentation skeleton; once deleted, they are no longer detected as fitted B\'ezier segments~\citep{cogan1961retinal,hammes2002pericytes,curtis2009microvascular,stitt2016progress,forster2021retinal,pramil2021macular,doi:10.1056/NEJMra1005073,broe2014retinal,popovic2019fractal}. \code{total_arc_length}, \code{branching_density}, and \code{mean_chord_length} therefore all decrease as DR severity rises, which accounts for the monotonically decreasing trend of ranks 1 and 4 and supplies the subtractive arm of rank 2.

\subsection{Chain 2: Caliber Biphasic Trajectory}

The second chain explains \code{mean_radius} (rank 3) through a biphasic time-course in which an early arteriolar dilation is overtaken by a late arteriolar narrowing. In early DR, two parallel processes widen the arteriolar lumen. First, impaired retinal autoregulation under sustained hyperglycemia produces tonic hyperperfusion. Second, mild local hypoxia and high glucose drive endothelial release of nitric oxide (NO); NO relaxes the arteriolar smooth muscle and produces functional vasodilation. Both arms push \code{mean_radius} upward in the early phase. In late DR, two further independent paths narrow the same vessels. \emph{Path A}: chronic oxidative stress thickens the vascular basement membrane (BM); BM thickening directly thickens the wall, reduces the effective diameter of the lumen, and stiffens the vessel. \emph{Path B}: cumulative downstream capillary closure (the late-stage end of Chain~1) removes the distal perfusion demand and the upstream arterioles atrophy in response. The two late-phase paths jointly produce a net reduction of \code{mean_radius}~\citep{stitt2016progress,roy2021retinal,doi:10.1056/NEJMra1005073,curtis2009microvascular,ashraf2021retinal,cheung2008retinal,klein2004relation,velayutham2020extended,xu2026quantitative}. Since the imaged DR cohort predominantly consists of patients beyond the early hyperperfusion window, the population-level Spearman correlation reflects the late-phase decline. However, the residual early-phase upward trend is preserved within the non-linear SHAP signatures of several caliber-related features (Appendix~\ref{app:track1-full}).

\subsection{Chain 3: Tortuosity Biphasic Trajectory and Late-Stage Neovessel Coupling}

The third chain explains \code{mean_tortuosity} (rank 5) and also resolves the apparent paradox of \code{branching_density} (rank 2) carrying a weak positive Spearman ($\rho = +0.014$, $p < 0.05$) together with a strongly negative tree-conditional \texttt{feat-SHAP corr} ($-0.55$). Tortuosity rises in both the early and the late phase of DR, but through different mechanisms.

\paragraph{Early-mid phase} 
 Four contributors push \code{mean_tortuosity} upward. (i) Pericyte loss, the same starting node as Chain~1, removes mural support from the endothelium, so the vessel wall is more compliant under pulsatile load. (ii) High axial shear stress from the early hyperperfusion arm of Chain~2 drives elongation and buckling of compliant segments. (iii) Impaired retinal autoregulation amplifies the pressure pulse delivered to the affected segments. (iv) Early basement-membrane remodeling alters the passive elastic ~\citep{roy2021retinal}. The four contributors all act in the early-mid disease window and form the early upward arm of the tortuosity time-course.

\paragraph{Late phase} 
 The end of Chain~1, that is, widespread vasoregression and cumulative capillary closure, produces local retinal hypoxia. Hypoxia sharply elevates vascular endothelial growth factor (VEGF), and VEGF drives neovessel proliferation. The new vessels are typically thin and short, so most of them fall below the AutoMorph segmentation and skeletonization thresholds and are not picked up as distinct fitted B\'ezier segments~\citep{aiello1994vascular,stitt2016progress,sasongko2011retinal,sasongko2016retinal,forster2021retinal,xu2026quantitative,fathimah2025retinal}. The consequences across the SHAP top-5 features are therefore asymmetric. \code{total_arc_length} and \code{mean_chord_length}, which sum or average over segments that are actually fitted, continue to decrease because the captured vessel network is dominated by the regressed mature tree rather than by the un-captured neovessels. However, \code{branching_density}, being more sensitive to the junction-level cues produced by neovessel sprouts at their basal anastomoses, picks up a faint additive signal. Furthermore, the characteristic geometry of neovascularization, with its tight loops and abrupt directional changes, reinforces \code{mean_tortuosity}. Thus, the mechanical remodeling of the early phase and the neovascularization of the late phase act in concert to produce the monotonically increasing trend observed in the cohort.

\paragraph{Summary mapping to SHAP top-5} 
In summary, Chain 1 accounts for the monotonically decreasing trend of ranks 1 and 4 (\code{total_arc_length}, \code{mean_chord_length}) and provides the subtractive component of rank 2 (\code{branching_density}). Chain 2 explains the late-phase decline in rank 3 (\code{mean_radius}), while its residual early-phase expansion remains detectable in the non-linearity audit of the full 20-feature set. Chain 3 drives the monotonically increasing trajectory of rank 5 (\code{mean_tortuosity}) through the synergistic superposition of early mechanical remodeling and late neovascularization. Furthermore, Chain 3 supplies the additive signal for \code{branching_density}; when combined with the subtractive influence of Chain 1, this creates a biphasic signature that reconciles the near-zero marginal Spearman correlation with the high non-linear feature importance.


\section{Datasets, Code, and Licenses}
\label{app:assetslink}

\subsection{Datasets}

\paragraph{Eyepacs, Aptos, Messidor Diabetic Retinopathy}
\url{https://www.kaggle.com/datasets/ascanipek/eyepacs-aptos-messidor-diabetic-retinopathy}. This dataset does not claim ownership of the original images. It is provided as a unified, resized, augmented, and preprocessed version for research and educational purposes.

\paragraph{EyePACS}
\url{https://www.kaggle.com/competitions/diabetic-retinopathy-detection}. 
License/terms: Custom Kaggle Competition Rules for the Diabetic Retinopathy Detection competition; dataset redistribution is not permitted.

\paragraph{APTOS-2019}
\url{https://www.kaggle.com/competitions/aptos2019-blindness-detection/rules#7-competition-data}. 
License/terms: Custom Kaggle Competition Rules; non-commercial academic/research and educational use only; redistribution is prohibited.

\paragraph{APTOS Gaussian Filtered}
\url{https://www.kaggle.com/datasets/sovitrath/diabetic-retinopathy-224x224-gaussian-filtered}. CC0: Public Domain.

\paragraph{Messidor-2}
\url{https://www.adcis.net/en/third-party/messidor2/}. 
License/terms: Custom research and educational use license; redistribution and unauthorized commercial use are prohibited. 
Acknowledgment: Kindly provided by the Messidor program partners 
(\url{https://www.adcis.net/en/third-party/messidor/}).

\paragraph{UK Biobank}
(\url{https://www.ukbiobank.ac.uk/}). Managed-access biomedical resource; usage is governed by the UK Biobank Access Procedures.

\subsection{Code and Models}

\paragraph{AutoMorph}
Source code for this pipeline is available from (\url{https://github.com/rmaphoh/AutoMorph}). Apache License
Version 2.0, January 2004

\paragraph{Stable Diffusion 2.1}
(\url{https://huggingface.co/sd2-community/stable-diffusion-2-1}). Openrail++ license.

\paragraph{LoRA adapter}
(\url{https://github.com/microsoft/LoRA}). MIT license.

\paragraph{ControlNet}
(\url{https://github.com/lllyasviel/ControlNet?tab=readme-ov-file}). Apache License Version 2.0, January 2004.

\paragraph{BiomedCLIP}
(\url{https://huggingface.co/microsoft/BiomedCLIP-PubMedBERT_256-vit_base_patch16_224}) MIT license.


\section*{NeurIPS Paper Checklist}

\begin{enumerate}

\item {\bf Claims}
    \item[] Question: Do the main claims made in the abstract and introduction accurately reflect the paper's contributions and scope?
    \item[] Answer: \answerYes{} 
    \item[] Justification: 
    Our three main claims (the B\'ezier Tree Encoding framework, parameter-level counterfactual verification, and vision--language decoupling) are explicitly stated in Introduction and directly supported by end-to-end evaluations in Experiments. Furthermore, the scope boundaries and limitations of our approach, including cohort-scale constraints and the focus on model-level rather than biological causality, are transparently discussed in Discussion.


\item {\bf Limitations}
    \item[] Question: Does the paper discuss the limitations of the work performed by the authors?
    \item[] Answer: \answerYes{}.
    \item[] Justification: The paper discusses limitations in the Discussion section, including the primarily computational nature of the validation, the need for prospective clinical follow-up, reduced statistical power from the strict reverse-counterfactual subset, and the small AD cohort with a high false-positive baseline. It also identifies parameter-native ControlNet conditioning as future work and reports computational cost in the Data and Reproducibility section.

\item {\bf Theory assumptions and proofs}
    \item[] Question: For each theoretical result, does the paper provide the full set of assumptions and a complete (and correct) proof?
    \item[] Answer: \answerNA{}.
    \item[] Justification: The paper does not present formal theoretical results, theorems, or lemmas requiring assumptions and proofs. The proposed framework is methodological and empirically validated through computational experiments.

\item {\bf Experimental result reproducibility}
    \item[] Question: Does the paper fully disclose all the information needed to reproduce the main experimental results of the paper to the extent that it affects the main claims and/or conclusions of the paper (regardless of whether the code and data are provided or not)?
    \item[] Answer: \answerYes{}.
    \item[] Justification: The paper discloses the key information required to reproduce the main experimental claims: dataset splits, model architectures, training hyperparameters, perturbation definitions, inference settings, evaluation metrics, and statistical tests. The appendix further reports implementation details for B\'ezier encoding, LoRA/ControlNet training, classifier training, and counterfactual scoring.

\item {\bf Open access to data and code}
    \item[] Question: Does the paper provide open access to the data and code, with sufficient instructions to faithfully reproduce the main experimental results, as described in supplemental material?
    \item[] Answer: \answerNo{}.
    \item[] Justification: The code is not publicly released at submission time, but will be made publicly available upon acceptance. The paper and supplemental material provide detailed experimental settings, data-processing procedures, and evaluation protocols to support reproducibility.

\item {\bf Experimental setting/details}
    \item[] Question: Does the paper specify all the training and test details (e.g., data splits, hyperparameters, how they were chosen, type of optimizer) necessary to understand the results?
    \item[] Answer: \answerYes{}.
    \item[] Justification: The paper specifies the experimental settings needed to understand the results, including data splits, preprocessing, model architectures, training hyperparameters, optimizer settings, perturbation design, inference configuration, evaluation metrics, and statistical testing. Additional training and implementation details are provided in the appendix and supplemental material.

\item {\bf Experiment statistical significance}
    \item[] Question: Does the paper report error bars suitably and correctly defined or other appropriate information about the statistical significance of the experiments?
    \item[] Answer: \answerYes{}.
    \item[] Justification: The paper reports appropriate statistical significance information for the main experimental claims, including significance tests for observational feature--disease associations and for counterfactual intervention effects relative to matched controls.

\item {\bf Experiments compute resources}
    \item[] Question: For each experiment, does the paper provide sufficient information on the computer resources (type of compute workers, memory, time of execution) needed to reproduce the experiments?
    \item[] Answer: \answerYes{}.
    \item[] Justification: The paper reports the compute resources used for the experiments in the Data and Reproducibility section, including GPU hardware, training/inference time, and the main software environment. These details provide the information needed to estimate the computational resources required to reproduce the reported experiments.
    
\item {\bf Code of ethics}
    \item[] Question: Does the research conducted in the paper conform, in every respect, with the NeurIPS Code of Ethics \url{https://neurips.cc/public/EthicsGuidelines}?
    \item[] Answer: \answerYes{}.
    \item[] Justification: The research conforms to the NeurIPS Code of Ethics. The study uses de-identified retinal imaging data and computational analyses, does not expose personally identifiable information, and does not make claims of direct clinical deployment. The paper also discusses limitations and the need for prospective clinical validation before any clinical use.

\item {\bf Broader impacts}
    \item[] Question: Does the paper discuss both potential positive societal impacts and negative societal impacts of the work performed?
    \item[] Answer: \answerYes{}.
    \item[] Justification: The paper discusses broader impacts, including the potential positive impact of interpretable counterfactual tools for biomedical hypothesis verification and retinal disease analysis. It also discusses possible negative impacts, including risks from over-interpreting computational counterfactuals, premature clinical use without validation, privacy concerns for medical imaging data, and fairness concerns across patient populations.
    
\item {\bf Safeguards}
    \item[] Question: Does the paper describe safeguards that have been put in place for responsible release of data or models that have a high risk for misuse (e.g., pre-trained language models, image generators, or scraped datasets)?
    \item[] Answer: \answerNA{}.
    \item[] Justification: The released components are task-specific scoring classifiers and a B\'ezier-conditioned vessel-rendering ControlNet whose output domain is constrained to retinal fundus geometries by the B\'ezier parametric hint; the model is not a general-purpose image generator and does not pose dual-use deepfake or scraping risks. No personally identifiable medical data are redistributed. 

\item {\bf Licenses for existing assets}
    \item[] Question: Are the creators or original owners of assets (e.g., code, data, models), used in the paper, properly credited and are the license and terms of use explicitly mentioned and properly respected?
   \item[] Answer: \answerYes{}.
    \item[] Justification: The paper cites all existing datasets, pretrained models, and software assets used, and summarizes their source, intended use, and license/terms of use in Appendix~\ref{app:assetslink}. No raw medical imaging data are redistributed.

\item {\bf New assets}
    \item[] Question: Are new assets introduced in the paper well documented and is the documentation provided alongside the assets?
    \item[] Answer: \answerNA{}.
    \item[] Justification: The paper does not release new datasets, model checkpoints, or other standalone assets at submission time. The code will be documented and publicly released upon acceptance, while no identifiable or raw medical imaging data will be redistributed. 

\item {\bf Crowdsourcing and research with human subjects}
    \item[] Question: For crowdsourcing experiments and research with human subjects, does the paper include the full text of instructions given to participants and screenshots, if applicable, as well as details about compensation (if any)? 
    \item[] Answer: \answerNA{}.
    \item[] Justification: The paper does not involve crowdsourcing experiments, newly recruited human subjects, participant instructions, or compensation. The study is based on existing de-identified retinal imaging datasets and computational experiments.

\item {\bf Institutional review board (IRB) approvals or equivalent for research with human subjects}
    \item[] Question: Does the paper describe potential risks incurred by study participants, whether such risks were disclosed to the subjects, and whether Institutional Review Board (IRB) approvals (or an equivalent approval/review based on the requirements of your country or institution) were obtained?
    \item[] Answer: \answerNA{}.
    \item[] Justification: The paper does not involve newly recruited human subjects, crowdsourcing, or prospective data collection. The study uses existing de-identified retinal imaging datasets for computational experiments, and no identifiable participant information is released.

\item {\bf Declaration of LLM usage}
    \item[] Question: Does the paper describe the usage of LLMs if it is an important, original, or non-standard component of the core methods in this research? Note that if the LLM is used only for writing, editing, or formatting purposes and does \emph{not} impact the core methodology, scientific rigor, or originality of the research, declaration is not required.
   \item[] Answer: \answerNA{}.
    \item[] Justification: The core method development in this paper does not involve LLMs as an important, original, or non-standard component. Any use of LLMs, if applicable, was limited to writing, editing, or formatting assistance and did not affect the methodology, experiments, or scientific claims.

\end{enumerate}
\end{document}